\documentstyle[times,psfig,graphics,astrobib,amssymb]{mn2e}
\newcommand{\be}{\begin{equation}}
\newcommand{\e}{\end{equation}}
\newcommand{\bear}{\begin{eqnarray}}
\newcommand{\ear}{\end{eqnarray}}
\newcommand{\nline}{\nonumber \\}
\newcommand{\f}{\frac}
\newcommand{\de}{{\rm d}}

\begin{document}

\title[Experimental constraints on reionization]
{Experimental Constraints on Self-consistent Reionization Models}
\author[Choudhury \& Ferrara]
{T. Roy Choudhury\thanks{E-mail: chou@sissa.it}
and
A. Ferrara\thanks{E-mail: ferrara@sissa.it}\\
SISSA/ISAS, via Beirut 2-4, 34014 Trieste, Italy}

\maketitle

\date{\today}

\begin{abstract}
A self-consistent formalism to jointly study cosmic
reionization and thermal history of the intergalactic medium (IGM)
in a $\Lambda$CDM cosmology is presented. The model implements most of
the relevant physics governing these processes, such as the
inhomogeneous IGM density distribution, three different classes of
ionizing photon sources (massive PopIII stars, PopII stars and QSOs),
and radiative feedback inhibiting star formation in low-mass
galaxies. By constraining the model free parameters with available
data on redshift evolution of Lyman-limit absorption systems,
Gunn-Peterson and electron scattering optical depths, Near InfraRed
Background (NIRB), and cosmic star formation history, we select a
fiducial model, whose main predictions are: {\it(i)} Hydrogen was
completely reionized at $z \approx 15$, while HeII
must have been reionized by $z \approx 12$, allowing for the
uncertainties in the ionizing photon efficiencies of stars. At
$z \approx 7$, HeIII suffered an almost complete recombination as a result of
the extinction of PopIII stars, as {\it required} by the
interpretation of the NIRB.  {\it(ii)} A QSO-induced complete HeII
reionization occurs at $z=3.5$; a similar double H reionization does
not take place due to the large number of photons with energies $> 1$
Ryd from PopII stars and QSOs, even after all PopIII stars have
disappeared. {\it(iii)} Following reionization, the temperature of the
IGM corresponding to the mean gas density, $T_0$, is boosted to
$1.5 \times 10^4$~K; following that it decreases with a relatively flat
trend. 
Observations of $T_0$ are consistent with the 
fact that He is singly ionized 
at $z \gtrsim 3.5$, while they are consistent with He being 
doubly ionized at $z \lesssim 3.5$. 
This might be
interpreted as a signature of (second) HeII reionization. {\it(iv)}
Only 0.3\% of the stars produced by $z=2$ need to be PopIII stars in
order to achieve the first hydrogen reionization. 
In addition, we get
useful constraints on the ionizing photon efficiencies 
(which are combination of the star-forming efficiency and 
the escape fraction of ionizing photons from collapsed haloes)
of PopII and
PopIII stars, namely, $\epsilon_{\rm PopII}< 0.01$, $0.002 <
\epsilon_{\rm PopIII} < 0.03$. 
Varying the efficiencies in these two
ranges does not affect the scenario described above.  Such model not
only relieves the tension between the Gunn-Peterson optical depth and
WMAP observations, but also accounts self-consistently for all known
observational constraints. We discuss how the results compare with
recent numerical reionization studies and other theoretical arguments.
\end{abstract}

\section{Introduction}

The study of cosmic reionization has witnessed a very strong advance
in the last few years (for a recent review see \citeNP{cf04}), thanks
both to the availability of QSOs located at redshifts $\ga 6$ which
allow to probe the physical state of the neutral part of the
intergalactic medium (IGM) via absorption line experiments
\cite{fwd++00,fnl+01,fnswbpr02,fss++03}, and to the determination of
the Thomson electron scattering optical depth $\tau_{\rm el}$ from the
first year WMAP data \cite{svp++03}.  A considerable tension from
these two data sets has arisen concerning the epoch of hydrogen
reionization. In brief, while the rapid increase in the
(Gunn-Peterson) Ly$\alpha$ opacity at $z\ga 5$ could be interpreted,
to a first analysis, as an indication of a reionization occurring at
$z \approx 6$, the large inferred values of $\tau_{\rm el}$ would
imply a reionization epoch at $z > 10$. 
In this regard one has to
recall that (i) the first year WMAP result has large error-bars; (ii) the limits 
on $\tau_{\rm el}$
depend on the exact analysis technique employed; 
(iii) the robustness of the result needs to be confirmed using more data, 
while on the 
other hand (i) the Gunn-Peterson test is very sensitive to even tiny
amounts of neutral hydrogen, resulting only in a lower limit to the
neutral hydrogen fraction, $x_{\rm HI} \ga 0.1$\% \cite{wbfs03}; (ii)
the experiment has been carried out on a handful of QSOs above
$z=5$ and therefore its statistical significance has still to be
evaluated; (iii) the abrupt increase might be in part due an incorrect
conversion of Ly$\beta$ to Ly$\alpha$ optical depths
\cite{songaila04}. Nevertheless, reasons for the apparent discrepancy
must be considered seriously.

A crucial issue about reionization is that this process is tightly
coupled to the properties and evolution of star-forming galaxies and
QSOs. Hence, the first requirement that any reionization model
should fulfill is that it should be able to reproduce the available
constraints concerning the luminous sources. Whereas a relatively
solid consensus has been reached on the luminosity function, spectra
and evolutionary properties of intermediate redshift QSOs, some
debate remains on the presence of yet undetected low-luminosity
QSOs powered by intermediate mass black holes at high redshift
\cite{ro04b}. The evolution of stellar radiation is instead much less
certain. Models corroborated by the observation of an excess in the
Near InfraRed Background (NIRB) in the wavelength range 1--4~$\mu$m
\cite{sf03,msf03,kagmm04,cbklm04} require that the first, metal-free
(PopIII) stars were very massive if they are to account for this
otherwise unexplained excess. The same data require that a quite rapid
transition from PopIII to ``normal'' PopII stars must have occurred at
about $z=9$, probably driven by an increase of the metallicity in the
active cosmic star formation sites above the critical value $Z_{\rm
crit} = 10^{-5\pm 1} Z_{\odot}$ \cite{sfno02,sfsob03}. Even
after that, predicting the joint reionization and star formation
histories self-consistently is not an easy task as mechanical and
radiative feedback processes can alter the hierarchical structure
formation sequence of the underlying dark matter distribution as far
as baryonic matter is concerned.

Hence our present aim is to build up self-consistent models that can
reconcile the discrepancy between the Gunn-Peterson optical depth and
the first year WMAP observations, and at the same time satisfy a
larger number of experimental requirements concerning the IGM
temperature evolution, He reionization, the number of Lyman-limit
systems, the NIRB excess interpretation, and the cosmic star formation
history. Although this might seem to a first sight a very ambitious
aim, it turns out that once the relevant physics is included, these
results allow a very clear scenario to emerge.

Nothing comes for free, unfortunately, and there is a price to pay for
this wealth of predictions that can be obtained via a relatively
simple, although physically rigorous, semi-analytical approach to this
problem. When compared to the numerical simulations, we can treat many
details of the reionization process only in an approximate manner (the
shape of ionized region around sources and their overlapping, just to
mention a few) and in terms of global averages (such as the filling
factor and the clumping factor of ionized regions). However, we have
to recall that in order to exploit the full power of the observational
data available to constrain models, one must be able to connect widely
differing spatial (and temporal) scales. In fact, it is necessary, at
the same time, to resolve the IGM inhomogeneities (sub-kpc physical
scales), follow the formation of the very first objects in which
PopIII stars form (kpc), the radiative transfer of their ionizing
radiation and the transport of their metals (tens of kpc), the build
up of various backgrounds (Mpc), and the effects of QSOs (tens of
Mpc). Thus, a proper modelling of the relevant physics on these
scales, which would enable a direct and {\it simultaneous} comparison
with all the available data set mentioned above, would require
numerical simulations with a dynamical range of more than 5 orders of
magnitude, which is far cry from the reach of our current
computational technology. To overcome the problem, simulations have
typically concentrated on trying to explain one (or few, in the best
cases) of the observational constraints. It is therefore difficult
from these studies to understand the extent to which their conclusions
do not conflict with a different set of experiments other than the one
they are considering.

The quest for a more general, data-supported scenario is the main
motivation for this study. Although our semi-analytical approach
provides a way to achieve this aim in a very straightforward manner
and, as we will see, with very satisfactory results, it is important
that its uncertainties are kept under control as much as possible. For
this reason we also compare its predictions with those from numerical
simulations focusing on a more restricted aspect of the general
picture. The final hope is that heavily data-constrained models, like
to one presented here, will help us to identify the unique way in
which cosmic reionization occurred.

The paper is organized as follows: In the next few sections, we
develop the analytical formalism for studying the ionization and
thermal history of the IGM, implementing most of the essential
physical processes. We discuss the evolution of the ionized regions in
Section 2, taking into account the inhomogeneous density distribution
of the IGM. The explicit form of the IGM density distribution used in
this work is discussed in Section 3.  The estimation of the number of
ionizing photons from different types of sources is given in Section
4. The formalism for studying the thermal and ionization histories of
different phases of the IGM is discussed in Section 5. We discuss some
properties of the absorption systems derived from our formalism in
Section 6.  Finally, we compare the model predictions with various
observations and discuss the effect of varying different free
parameters in Section 7. The final section presents the summary of our
main results and compares the model with other semi-analytical models
and numerical simulations.

\section{Evolution of ionized regions}

In the following sections, we shall highlight the crucial points of
our formalism and try to develop a consistent picture of cosmic
reionization which derives from all known experimental constraints on
such a process. Such formalism allows us to track (i) the evolution of
the volume filling factors of ionized hydrogen and helium regions and
(ii) the thermal and ionization evolution in {\it each} of these
regions separately and self-consistently.

\subsection{Volume filling factor of ionized regions}

Let us first start with individual ionized regions: these can either
be singly-ionized hydrogen (HII) or doubly-ionized helium (HeIII). We
do not consider the analogous HeII regions, as typically these match
the HII ones and, in any case, very sparse observational data have
been collected concerning this ionization state of the IGM. To a first
approximation, it is usually assumed that the IGM can be described as
an uniform medium with small scale clumpiness taken into account
through a so-called ``clumping factor''. In such cases, the evolution of
the volume filling factor $Q_{\rm HII}$ for the HII regions will be
described by \cite{sg87}
\be
\f{\de Q_{\rm HII}}{\de t} 
= \f{\dot{n}_{\rm ph}}{n_H} 
- Q_{\rm HII} {\cal C}_{\rm HII} \f{n_e}{a^3} \alpha_R
\label{eq:dqi_dt}
\e
where $\dot{n}_{\rm ph}$ is the rate of ionizing photons per unit
volume, $n_H$ is the hydrogen atoms number density, ${\cal C}_{\rm
HII}$ is the clumping factor for the ionized regions, $a$ is the
expansion factor, and $\alpha_R$ is the recombination coefficient.
Under the assumption of homogeneity, the same equation holds for the
ionized hydrogen mass fraction as well. One can write a similar
equation for the volume filling factor $Q_{\rm HeIII}$ of HeIII
regions too. In this picture the reionization is said to be complete
when individual ionized regions overlap, i.e., $Q_i = 1$. The above
equation can be solved given a model for the source term $\dot{n}_{\rm
ph}$, and a value for the clumping factor ${\cal C}_i$.  There are few
points worth noting: (i) The quantity (corresponding to the
recombination term on the right hand side) $Q_{\rm HII} {\cal C}_{\rm
HII} n_e n_H \alpha_R$ gives the average recombinations per unit time
per unit volume in the universe.  (ii) Also note that we have
implicitly assumed that each photon is absorbed shortly after it is
emitted, i.e., its mean free path is much smaller than the horizon
size (which is true for $z > 2$). In case the photon is able to travel
large distances before it is absorbed, one has to take into account
the fact that photons might be redshifted below the ionization edge of
hydrogen.  (iii) Finally, one should remember that the quantity
$\dot{n}_{\rm ph}$ takes into account only those photons which escape
into the IGM. The number of photons produced by the source can be much
larger; however, a large fraction of those photons will be absorbed in
ionizing the high density halo surrounding the luminous source.

\subsection{Inhomogeneous Reionization}

In the above picture, the inhomogeneities in the IGM are considered
simply in terms of the clumping factor in the effective recombination
rate without taking into account the density distribution of the IGM.
The importance of using a density distribution of the IGM lies in the
fact that regions of lower densities will be ionized first, and
high-density regions will remain neutral for a longer time. The main
reason for this is that the recombination rate is higher in
high-density regions where dense gas becomes neutral very quickly. Of
course, there will be a dependence on how far the high density region
is from an ionizing source. A dense region which is very close to an
ionizing source will be ionized quite early compared to, say, a
low-density region which is far away from luminous sources.

The first effort in addressing such issues were carried out by
\citeN{mhr00} (MHR, hereafter). To summarize the main point of this approach,
consider, first, the situation where all the individual ionized
regions have overlapped (the so-called post-overlap stage;
\citeNP{gnedin00}). In this scenario, all the low-density regions of
the universe will be highly ionized, while there will be some high
density peaks (like the collapsed systems) which will still remain
neutral. Thus, as a reasonable first approximation (MHR), we assume
that all regions with overdensities $\Delta < \Delta_i$ are ionized
(the index $i$ refers to the different ionized species), while regions
with $\Delta > \Delta_i$ are completely neutral, with $\Delta_i$
increasing as time progresses (i.e., more and more high density
regions are getting ionized). We shall discuss the equation governing
the evolution of $\Delta_i$ later in this section, while the results
for different model parameters will be discussed in Section 7 (see,
for example, Fig. 1).  Note that, according to this scenario, the
reionization is defined to be completed when all the regions with
$\Delta < \Delta_i$ are ionized -- one does not need to ionize the
whole IGM to complete the reionization process. The effect of this
assumption is that only the low-density regions will contribute to the
clumping factor (regions whose density is above $\Delta_i$ are assumed
to be neutral, hence they need not be taken into account while
calculating the clumping factor).

The situation is slightly more complicated when the ionized regions
are in the pre-overlap stage. At this stage, it is assumed that a
volume fraction $1 - Q_i$ of the universe is completely neutral
(irrespective of the density), while the remaining $Q_i$ fraction of
the volume is occupied by ionized regions. However, within this
ionized volume, the high density regions (with $\Delta > \Delta_i$)
will still be neutral. Once $Q_i$ becomes unity, all regions with
$\Delta < \Delta_i$ are ionized and the rest are neutral. The
high-density neutral regions manifest themselves as the Lyman-limit
systems in the QSO absorption spectra. The reionization process after
this stage is characterized by increase in $\Delta_i$ -- implying that
higher density regions are getting ionized gradually.

To develop the equations embedding the above physical picture, we need
to know the probability distribution function $P(\Delta) \de \Delta$
for the overdensities. We shall discuss the form of $P(\Delta)$ in the
next section, but given a $P(\Delta) \de \Delta$, it is clear that
only a mass fraction
\be
F_M(\Delta_i) = \int_0^{\Delta_i} \de \Delta ~ \Delta ~ P(\Delta)
\e
needs be ionized, while the remaining high density regions will be
completely neutral because of high recombination rates. The
generalization of equation (\ref{eq:dqi_dt}), appropriate for this
description is given by MHR (see also \citeNP{wl03})
\be
\f{\de [Q_{\rm HII} F_M(\Delta_{\rm HII})]}{\de t} = 
\f{\dot{n}_{\rm ph}(z)}{n_H} 
- Q_{\rm HII} \f{\alpha_R(T) n_e R(\Delta_{\rm HII})}{a^3}
\label{eq:qifm}
\e
where $Q_{\rm HII} \alpha_R(T) n_{\rm HII} n_e R(\Delta_{\rm HII})$
gives the number of recombinations per unit time and volume. The
factor $R(\Delta_{\rm HII})$ is the analogous of the clumping factor,
and is given by
\be
R(\Delta_{\rm HII}) = \int_0^{\Delta_{\rm HII}} \de \Delta ~ 
\Delta^2 ~ P(\Delta)
\e
The reionization is complete when $Q_{\rm HII} = 1$; at this point a
mass fraction $F_M(\Delta_{\rm HII})$ is ionized, while the rest is
(almost) completely neutral.

Note that there are two unknowns $Q_{\rm HII}$ and $F_M(\Delta_{\rm
HII})$ in equation (\ref{eq:qifm}) [and similarly for the HeIII
regions]. For the post-overlap stage, we put $Q_i=1$, and can solve
the equation for $\Delta_i$. However, for the pre-overlap stage, we
have to deal with both the unknown and it is thus impossible to solve
it without more assumptions. One can fix either $\Delta_i$ or $F_M$
(the ionized mass fraction). There is no obvious way of dealing with
this problem. In this work, we assume that $\Delta_i$ does not evolve
with time in the pre-overlap stage, i.e., it is equal to a critical
value $\Delta_c$. To fix this $\Delta_c$, we note that results do not
vary considerably as $\Delta_c$ is varied from $\sim 10$ to $\sim
100$. For definiteness, we take the value corresponding to the typical
overdensity of collapsed haloes at the virial radius. The exact value
of this overdensity depends on the density profile of the halo; it can
be shown that it is $\approx 59.2$ for a isothermal profile, while it
is $\approx 63.7$ for the NFW profile. In this paper, we can safely
assume $\Delta_c = 60$, and also assume that this critical value is
the same for HII and HeIII. Once $\Delta_i$ is fixed, one can follow
the evolution of $Q_i$ until it becomes unity. Following that, we
enter the post-overlap stage, where the situation is well-described by
equation (\ref{eq:dqi_dt}).

\section{Probability distribution $P(\Delta)$}

There can be various approaches to determine the density distribution
of the IGM at various redshifts. It is clear that various complicated
physical processes will not allow us to obtain a simple distribution
from analytical calculations. One has to resort to either simulations,
or some sort of approximation schemes.  In the approach followed by
MHR, one uses a density distribution inspired by hydrodynamical
simulations. While such approaches are widely used, one should keep in
mind that the limitations related to box size and resolution are
inherent to every simulation. On the other hand, there are approaches
where the density distribution is obtained from some approximation
scheme. These approximation schemes seem to be reasonable in the
linear or quasi-linear regime of density fluctuations, while they are
more inaccurate when non-linear effects creep in.

In this work, we shall use such an approximation to describe the
low-density IGM. There are several reasons to believe that the
low-density regions of the IGM are well described by the lognormal
distribution (see, for example, \citeNP{cps01}; \citeNP{csp01}; 
\citeNP{vmht02}), which
is shown to be in excellent agreement with numerical simulations
\cite{bd97}.  However, the lognormal distribution seems to fail at
very high densities (say, when the densities are typical to that of
collapsed haloes). To see this, note that at high densities ($\Delta \gg 1$), 
the
lognormal distribution has the limiting form $P(\Delta) \sim
\Delta^{-\ln \Delta/\sigma^2 - 1}$, whereas it is expected that it
should follow a simple power-law $P(\Delta) \sim \Delta^{\beta}$. In
fact, if dark matter haloes have a density profile of the form
$r^{3/(\beta + 1)}$, then it can be shown that $P(\Delta)$ should fall
as $\Delta^{\beta}$. For an isothermal profile, one has $\beta =
-2.5$, while for a NFW profile, the value of $\beta$ varies from $-4$
in the center of the halo to $-2$ at the virial radius.

We thus assume the probability distribution of the overdensities to be
lognormal at low densities, changing to a power law form at high
densities:
\bear
P(\Delta) &=& \f{A}{\sigma \Delta \sqrt{2 \pi}}
{\rm e}^{-\f{1}{2 \sigma^2} 
\left(\ln \Delta - \mu\right)^2}   ~~~~~~~~~~~~~~~~~~~~~~{\rm if}~\Delta < \Delta_V
\nline
&=& \f{A}{\sigma \Delta_V \sqrt{2 \pi}}
{\rm e}^{-\f{1}{2 \sigma^2} 
\left(\ln \Delta_V - \mu\right)^2} 
\left(\f{\Delta}{\Delta_V}\right)^{\beta} {\rm if}~\Delta > \Delta_V
\nline
\ear
where $\sigma$ is rms linear mass fluctuations in
baryons.\footnote{Throughout this paper we will assume a flat universe
with total matter, vacuum, and baryonic densities in units of the
critical density of $\Omega_m = 0.27$, $\Omega_{\Lambda} = 0.73$, and
$\Omega_b h^2 = 0.024$, respectively, and a Hubble constant of $H_0 =
100\,h$ km s$^{-1}$ Mpc$^{-1}$, with $h=0.72$.  The parameters
defining the linear dark matter power spectrum are $\sigma_8=0.9$,
$n=0.99$, $\de n/\de \ln k =0$.}  If we want the derivative of
$P(\Delta)$ to be continuous at the transition overdensity $\Delta_V$,
then it follows that the slope $\beta$ and $\Delta_V$ must be related
by
\be
-1 -\f{1}{\sigma^2} (\ln \Delta_V - \mu) = \beta
\e
The parameters $A$ and $\mu$ are determined by normalizing the volume
and mass to unity. In most cases, the power law form of the
probability distribution comes into effect only for the post-overlap
phase.
 
In general, haloes could have $\beta$ varying with the distance from
center depending on the density profile.  Simulations also suggest a
redshift evolution of $\beta$, with $\beta = -2.5$ at high redshifts,
while it is $\approx -2.2$ at redshifts around 2 \cite{mcor96}. In
this work, however, we assume a constant value of $-2.3$ independent
of the redshift and the distance from the center. This is reasonable
as long as we are not too close to the center of the halo; also most
of the results are insensitive to $\beta$ as long it is within the
range $[-2.2, -2.5]$). Given a value of $\beta$, the value of
$\Delta_V$ will, in general, evolve with time.

To proceed further in the solution of equation (\ref{eq:qifm}), one
has to estimate two quantities: (i) the photon production rate
$\dot{n}_{\rm ph}(z)$, and (ii) the temperature, $T$, of the ionized
regions. We discuss the method adopted to obtain these estimates in
the next sections.

\section{Photon production rate}

\subsection{Photons from galaxies}

One can use the Press-Schechter formalism to estimate the mass
function of dark matter haloes of mass $M$ which collapsed at a
redshift $z_c$. In this paper, we shall use the formalism of
\citeN{sasaki94} for calculating the formation and merging rates of
dark matter haloes.  Assuming a model where the star formation rate (SFR) 
peaks around the
dynamical time-scale of the halo, it will form stars at the rate
\cite{co00,cs02}
\be
\dot{M}_{\rm SF}(M,z,z_c) = \epsilon_{\rm SF}
\left(\f{\Omega_B}{\Omega_m} M \right) 
\f{t(z)-t(z_c)}{t_{\rm dyn}^2(z_c)} 
{\rm e}^{-\f{t(z)-t(z_c)}{ t_{\rm dyn}(z_c)}}
\e
where $\epsilon_{\rm SF}$ is the efficiency of star formation. We can then write the cosmic SFR per unit comoving volume at a redshift $z$,
\be
\dot{\rho}_{\rm SF}(z) = \int_z^{\infty} \de  z_c 
\int_{M_{\rm min}(z_c)}^{\infty} \de  M \dot{M}_{\rm SF}(M,z,z_c) N(M,z,z_c),
\label{eq:sfr}
\e
where $N(M,z,z_c) \de  M \de  z_c$  gives the number of haloes within a mass range $(M, M + \de M)$ formed within a redshift interval $(z_c, z_c + \de z_c)$ and surviving
down to redshift $z$. The lower integration limit,  $M_{\rm min}(z_c)$, takes into account the fact that low mass haloes will not be able to cool and form stars. Also, note that
we do not take into account possible disruption of star forming haloes due to energy injection from exploding supernovae. We shall discuss the choice of  $M_{\rm min}(z_c)$
in detail in a later section.

Putting all the relevant expressions together, one can write the cosmic SFR in a neater form:
\be
\dot{\rho}_{\rm SF}(z) = \bar{\rho}_B
\f{1}{D(z)}\int_z^{\infty} 
\de  z_c \epsilon_{\rm SF}  F_1(z,z_c) {\cal I} (z_c)
\e
where
\be
F_1(z,z_c) = 
\left[\f{\dot{D}(z_c)}{D(z_c) H(z_c)}\right] 
\f{D(z_c)}{(1 + z_c)} \f{t(z)-t(z_c)}{t_{\rm dyn}^2(z_c)} 
{\rm e}^{-\f{t(z)-t(z_c)}{t_{\rm dyn}(z_c)}}
\e
and
\be
{\cal I} (z_c) = \int_{\nu(M_{\rm min}[z_c])}^{\infty} 
\de  \nu 
\left(\sqrt{\f{2}{\pi}} {\rm e}^{-\nu^2/2} \right)
\nu^2
\e
In the above expressions, $D(z)$ gives the linear growth factor of dark matter perturbations, and $\nu(M) = \delta_c/[D(z) \sigma_{\rm DM}(M)]$. In this 
paper, we fix the critical overdensity for collapse, $\delta_c$, to 1.69.

The rate of ionizing photons per unit volume per unit frequency range would be
\be
\dot{n}_{\nu, G}(z) = \left[\f{\de N_{\nu}}{\de M}\right]
\bar{\rho}_B
\f{1}{D(z)}\int_z^{\infty} 
\de  z_c \epsilon_{\rm SF}
f_{\rm esc}
F_1(z,z_c) {\cal I} (z_c)
\e
where $f_{\rm esc}$ is the escape fraction of ionizing photons from the star forming haloes, and $\de N_{\nu}/\de M$ gives the number of photons emitted per  
frequency range per unit mass of stars. Given the spectra of stars of different masses in a galaxy, and their Initial Mass Function (IMF), this quantity can be 
computed in a straightforward way. The IMF and spectra will depend on the details of star formation and metallicity, and can be quite different for Pop II and 
Pop III stars. We shall return to this point at the end of this section. Given the above quantity, it is straightforward to calculate the total number density of ionizing photons 
emitted at a particular frequency range $[\nu_{\rm min}, \nu_{\rm max}]$ by galaxies per unit time.

\subsection{Ionizing photons from PopII/PopIII stars}

While considering the number of photons from stars, one has to keep in mind that the nature of stars which were formed early could be very different 
from what we observe at lower redshifts. The physical motivation behind this assumption is the commonly accepted view that early stars are metal-free 
and hence the IMF would be top-heavy, i.e., dominated by high-mass stars \cite{sfno02}. Currently, the observational 
support for this assumption comes mainly from the analysis of the 
cosmic NIRB data  \cite{sf03,msf03}. Hence, in this work we 
consider the possibility that at early redshifts, a population of metal-free, massive (PopIII) stars produced a large number of photons. On the other hand, 
star formation at low redshifts, as we know, is dominated by high metallicity PopII stars with the usual Salpeter-like IMF. 
It is believed that the transition from PopIII to PopII stars occurred because 
of an increase in the metallicity in the active star-forming regions above a
critical value of $Z_{\rm crit} = 10^{-5\pm1} Z_{\odot}$.  
However, it is not clear how abrupt the transition has been, due to the poorly
known amplitude of the optical background at 0.6-0.8 $\mu$m. 
We shall therefore denote this transition redshift by $z_{\rm trans}$, and study the effects of its 
variation. 
We should mention that PopIII star formation may continue at lower redshifts
with a decreasing rate, however the interpretation of the NIRB
requires that their contribution to the 
ionizing flux be negligible with respect to the PopII stars. Although some
metal pollution has to take place for transition to PopII stars, 
it is necessary that the majority of the NIRB-contributing PopIII
stars have masses such that a black hole is the final product of their
evolution \cite{sf03}.

From the above, it thus turns out that the star formation is made up of two components:
\be
\dot{n}_{{\rm ph}, G}(z) = \dot{n}_{\rm ph, PopII}(z) + \dot{n}_{\rm ph, PopIII}(z);
\e
the previous expression involves two free parameters which are the 
ionizing photon efficiencies of the PopII and PopIII stars, namely:
$\epsilon_{\rm Pop II} \equiv \epsilon_{\rm SF, PopII} f_{\rm esc, PopII}$ and $\epsilon_{\rm Pop III} \equiv \epsilon_{\rm SF, PopIII} 
f_{\rm esc, PopIII}$. We assume that stellar population within the galaxies formed at $z > z_{\rm trans}$ is dominated by PopIII stars, while at lower 
redshifts it is dominated by PopII stars. Since the SFR peaks at the dynamical time scale after the formation of the halo and decreases exponentially thereafter, the formation rate 
of  PopIII stars continues for some amount of time after $z_{\rm trans}$, and hence the transition is not instantaneous. By this assumption, we are implicitly 
neglecting the effects of galactic self-enrichment on the IMF transition. Although locally (i.e., inside a given galaxy), the transition from PopIII to PopII star 
formation mode can occur at epochs different (either earlier or later) than $z_{\rm trans}$, the studies of NIRB data set the epoch when the bulk of the cosmic stars became PopII at $z_{\rm trans}$.

To calculate the number of photons produced per unit mass  of PopII star formed $[\de N_{\nu}/\de M]_{\rm PopII}$, we use the stellar spectra calculated using STARBURST99 \footnote{http://www.stsci.edu/science/starburst99}
\cite{lsg++99}, with metallicity $Z=0.001=0.05 Z_{\odot}$ 
and standard Salpeter IMF. 
The number of photons per unit mass of star formed, when integrated over appropriate
frequencies, gives $(8.05, 2.62, 0.01) \times 10^{60} M_{\odot}^{-1}$ for 
(HII, HeII, HeIII) respectively. 
For calculating $[\de N_{\nu}/\de M]_{\rm PopIII}$, we first assume that the IMF of the PopIII stars is dominated by very high mass stars. Since the number of photons produced per unit mass of stars formed is somewhat independent of the stellar mass for high mass stars, it is independent of the precise shape of the IMF (as long as the IMF is dominated by high mass stars). The value of $[\de N_{\nu}/\de M]_{\rm PopIII}$ is calculated using stellar spectra for high mass stars ($\geq 300 M_{\odot}$) as given by \citeN{schaerer02}.
When integrated over appropriate frequencies, this quantity is equal to
$(2.69, 4.46, 1.36) \times 10^{61} M_{\odot}^{-1}$ for 
(HII, HeII, HeIII) respectively. 

\subsection{The minimum mass for the star-forming haloes}

The quantity $M_{\rm min}(z_c)$ in equation (\ref{eq:sfr}) depends on the cooling efficiency of haloes. We assume that molecular cooling is active at high redshifts, and 
$M_{\rm min}(z_c)$ increases with decreasing $z$ \cite{fc00}. 
However, it is also possible that the molecular cooling within the mini-haloes 
could be highly suppressed due to photo-dissociation of hydrogen molecules. 
We shall study the effect of  mini-halo suppression in Section 7.5.
At lower redshifts (say $z < 10$), when the mass function of collapsed haloes 
is dominated by relatively high-mass haloes, the SFR is more or less independent of whether molecular cooling is effective or not.

There is a further factor which needs to be taken into account -- the radiative feedback from stars. Once the first galaxies form stars, their radiation
will ionize and heat the surrounding medium, increasing the mass scale (often referred to as the {\it filtering mass}) above which baryons can collapse 
into haloes within those regions. The minimum mass of haloes which are able to cool is thus much higher in ionized regions than in neutral ones. 
Since we are considering a multi-phase IGM, one needs to take into account the heating of the ionized regions from the beginning (even before the 
actual overlap has started). As we compute the temperature of the ionized region self-consistently, we can calculate the minimum circular 
velocity of haloes that are able to cool using the relation:
\be
v_c^2 = \f{2 k_{\rm boltz} T}{\mu m_p}
\label{radfeed}
\e
where $\mu$ is the mean molecular weight: $\mu m_p = \rho_b/n_b$. 
Typically, for a temperature of $3 \times 10^4 K$, the minimum circular velocity is $\sim$ 30--50 km s$^{-1}$, which is comparable to the filtering scale 
obtained from numerical simulations of \citeN{gnedin00} after the universe
has reionized.

Note that the above value of $v_c$ will evolve according to the temperature 
of the gas. 
An alternative feedback prescription is to fix the
minimum circular velocity of haloes allowed to form stars at a given value, which can be taken to be in the range $v_c=35$--50~km s$^{-1}$
\cite{gnedin00,ksui01}. 
We shall explore the effect of such different feedback prescriptions 
later in Section 7.

\subsection{Photons from QSOs}

In order to calculate the number of ionizing photons from QSOs,  
we shall follow the simple formalisms developed by \citeN{wl03} and 
\citeN{mds04}. The only difference in our approach is that we use 
a different prescription for calculating the merging and formation 
of dark matter haloes. In the previous works, the merging rates 
of dark matter haloes were based on the formalism by \citeN{lc93}, while 
our model uses the \citeN{sasaki94} methodology. For the reason
of completeness, we include the details of the model for calculating the 
luminosity function of QSOs in this section.

The key assumption to calculate the luminosity of QSOs is that the mass of the accreting black hole $M_{\rm bh}$ is correlated with the circular velocity 
$v_c$ of the collapsed halo through the relation:
\be
M_{\rm bh} = \epsilon v_c^{\alpha}
\e
It is argued that $\alpha = 5$ in a self-regulated growth of super-massive black holes \cite{sr98}, a value found to match observations of local universe.

It is then reasonable to assume that the black hole radiates with the Eddington luminosity (in the B-band), given by
\be
\f{L_B}{L_{\odot,B}} = 5.7 \times 10^3 \f{M_{\rm bh}}{M_{\odot}}
\e
which gives
\be
\f{L_B}{L_{\odot,B}} = \beta \left(\f{M}{M_{\odot}}\right)^{\alpha/3}
\label{eq:lb_m}
\e
where we have used the relation between the circular velocity and 
the mass of a virialised halo \cite{cp02} to obtain
\bear
\beta &=& 5.7 \times 10^3 \epsilon_0 10^{-4 \alpha}
\left(\f{H_0^2 18 \pi^2}{H^2(z) \Delta_{\rm vir}(z)}\right)^{\alpha/6}
\nline
&\times& \left[1-
\f{2 \Omega_{\Lambda} H_0^2}
{3 H^2(z) \Delta_{\rm vir}(z)} \right]^{\alpha} h^{\alpha/3}
\ear
with
\be
\epsilon_0 = \f{\epsilon (159.4 {~\rm km~s^{-1}})^{\alpha}}{M_{\odot}}
\e
Note that, even if the black hole
radiates in a sub-Eddington rate, the corresponding uncertainty
can be absorbed into the value of $\beta$.
The luminosity function of QSOs (i.e., the number of QSOs per unit comoving volume per unit luminosity range) will be given by
\be
\psi(L_B, z) \de L_B = \int_{\infty}^z \de z_c ~ N(M,z,z_c) \de M
\e
where $M$ and $L_B$ are related by equation (\ref{eq:lb_m}). If we assume that each QSO lives for a time $t_{\rm qso}(z) \ll (\dot{a}/a)^{-1}$, then the QSO activity 
can be taken to be almost instantaneous and we can approximate
\be
\int_{\infty}^z \de z_c ~ N(M,z,z_c) \approx
\f{\de z}{\de t} t_{\rm qso}(z) N(M,z,z)
\e
Then
\bear
\psi(L_B, z) &=& \f{\de z}{\de t} t_{\rm qso}(z) N(M,z,z) \f{\de M}{\de L_B}
\nline
&=& \f{3}{\alpha L_B} 
M N_M(z) \nu^2 \left[\f{\dot{D}(z)}{D(z) H(z)}\right] H(z) t_{\rm qso}(z)
\nline
\label{eq:psilb_main}
\ear
We fix $t_{\rm qso}(z) = 0.035 t_{\rm dyn}(z)$ \cite{mds04}. Given $\alpha = 5$, one can fix the free parameter $\epsilon_0$ by comparing the model 
with observations of QSO luminosity function.

This simple procedure works very well at intermediate and high redshifts, but fails to match the observations 
at low redshifts. One can introduce a phenomenological function in equation (\ref{eq:psilb_main}) to take into account the break in the luminosity function at high luminosities. 
We find that a modified $\psi(L_B, z)$ of the form \cite{mds04}
\be
\psi(L_B, z) \to \psi(L_B, z) \exp\left[-\f{M}{10^{11.25+z} M_{\odot}}\right]
\e
is good enough in matching the low redshift observations. This cutoff is at very high luminosities and has virtually no effect at $z > 3$.

Given the luminosity function, the rate of ionizing photons from QSOs per unit volume per unit frequency range will be
\be
\dot{n}_{\nu, Q}(z) = \int_0^{\infty} \de L_B \psi(L_B,z) 
\f{L_{\nu}(L_B)}{h_P \nu}
\e
We next use the mean UV QSO spectrum \cite{sb03}
\be
\f{L_{\nu}(L_B)}{{\rm ergs~s}^{-1} {\rm Hz}^{-1}} = \f{L_B}{L_{\odot,B}} 
10^{18.05} \left(\f{\nu}{\nu_H}\right)^{-1.57}
\e
which then gives,
\bear
\dot{n}_{\nu, Q}(z) &=& 
\left[\f{10^{18.05} {\rm ergs~s}^{-1} {\rm Hz}^{-1}}{L_{\odot,B}}\right]
\f{1}{h_P \nu_H} \left(\f{\nu}{\nu_H}\right)^{-2.57}
\nline
&\times&
\int_0^{\infty} \de L_B ~ L_B ~ \psi(L_B,z) 
\ear
The rate of ionizing photons from QSOs is obtained simply by integrating over all relevant frequencies.

This simple phenomenological model suits very nicely the purposes of this paper.  Alternatively, one can simply use the observed luminosity function for QSOs for 
calculating the number of ionizing photons. Since the above model matches the perfectly well with observations at low redshifts, none of the results would 
be affected. At high redshifts, say $z > 6$,  the contribution from QSOs are negligible compared to that of galaxies, and can be ignored.

\section{Thermal evolution}

In the previous sections, we have discussed the evolution of the ionized volume filling factors, and various physical quantities related to them. 
It is clear that the baryonic universe will behave as three-phase medium constituted by: (i) completely neutral regions, (ii) regions where hydrogen is 
ionized and helium is singly ionized, and (iii) a region where both species are fully ionized. We have assumed that the ionization front for the 
doubly-ionized helium can never overtake that for ionized hydrogen -- which is found to be {\it always} true for the type of spectra we are 
using for the ionizing sources (note that a much harder spectrum can always make the ionization front for the doubly-ionized helium leading the 
ionized hydrogen region). 

The thermal evolution equations are solved separately for each of the three regions. 
In the absence of heating sources 
the evolution is nearly trivial for the neutral region, 
with the temperature decreasing adiabatically. 
However, there is always a background of hard photons which can heat the neutral IGM. 
Since the temperature of the neutral region hardly affects the reionization history, we 
ignore such hard photons and let the temperature of the neutral regions decrease adiabatically.
In the ionized regions, 
the temperature can be calculated using the dynamical equation
\be
\f{\de T}{\de t} = -2 H(z) T -\f{T}{\sum_i X_i} 
\f{\de \sum_i X_i}{\de t}
+ \f{2}{3 k_{\rm boltz} n_B (1 + z)^3} \f{\de E}{\de t}
\label{eq:dTdt}
\e
where
\be
X_i \equiv \f{n_i m_p}{\bar{\rho}_B}
\e
and $\de E/\de t$ gives the net heating rate per baryon. For most purposes, it is sufficient to take into account the photoionization heating 
and recombination cooling (and Compton cooling off CMB photons, 
which can be important at 
high redshifts). 
For example, in the regions where only hydrogen is ionized, we have
\bear
\f{\de E}{\de t} &=& 
n_{\rm HI} (1 + z)^6 \int_{\nu_H}^{\infty} \de \nu ~ 
\lambda_H(z;\nu) \f{\dot{n}_{\nu}(z)}{Q_{\rm HII}(z)} 
\sigma_H(\nu) h_P(\nu - \nu_H)
\nline
&-& R(\Delta_{\rm HII}) \alpha_{RC}(T) n_{\rm HII} n_e (1 + z)^6
\label{eq:dEdt}
\ear
where $\lambda_H(z;\nu)$ is the {\it proper} mean free path for hydrogen ionizing photons with frequency $\nu > \nu_H$.
It is found from observations at low redshifts that  
$\lambda_H(z;\nu) \propto \nu^{1.5}$; this is understood 
from the frequency-dependence of the absorption cross section 
$\sigma_H(\nu) \propto \nu^{-3}$, and the column density distribution of QSO absorption lines 
$\de N/\de N_{\rm HI} \propto N_{\rm HI}^{-3/2}$ \cite{pwrcl93}. We assume this relation to be valid for all redshifts. Note that this frequency dependence of the mean free path hardens the ionizing spectrum. However, at frequencies below the ionization threshold
of HeII, the diffuse recombination radiation from the IGM tends to compensate for this
hardening \cite{hm96}. Given this, we assume $\lambda_H(z;\nu) = \lambda_{H,0}(z)$ for
$\nu \leq \nu_{\rm HeII}$ and $\lambda_H(z;\nu) = \lambda_{H,0}(z) (\nu/\nu_{\rm HeII})^{1.5}$ for
$\nu > \nu_{\rm HeII}$.
The procedure for calculating $\lambda_{H,0}(z) \equiv \lambda_H(z;\nu=\nu_H)$ is described in the next section.

The equation for evolution of the temperature has to be supplemented by those for the ionization of the individual species. In the most general case, 
one has three independent species $X_i = \{X_{\rm HI}, X_{\rm HeI}, X_{\rm HeIII}\}$, with other species being given by 
\bear
X_{\rm HII} &=& 1 - Y - X_{\rm HI}; ~~ 
\nline
X_{\rm HeII} &=& \f{Y}{4} - X_{\rm HeI} - X_{\rm HeIII}; ~~ 
\nline
X_e & =& X_{\rm HII} + X_{\rm HeII} + 2 X_{\rm HeIII}
\ear
where $Y = 0.24$ is the helium weight fraction.
For example, the evolution of $X_{\rm HI}$ in the HII region is given by
\bear
\f{\de X_{\rm HI}}{\de t} &=& - X_{\rm HI} (1 + z)^3 
\int_{\nu_H}^{\infty} \de \nu ~ \lambda_H(z;\nu) 
\f{\dot{n}_{\nu}(z)}{Q_{\rm HII}(z)} \sigma_H(\nu)
\nline 
&+& 
R(\Delta_{\rm HII}) \alpha_R(T) X_{\rm HII} X_e \f{\bar{\rho}_B}{m_p} (1 + z)^3
\label{eq:dXHIdt}
\ear
Similar equations, though slightly more complicated, can be written down for other regions too.

In passing, note that the volume emissivity is given by
\be
\epsilon_{\nu}(z) = \dot{n}_{\nu}(z) h_P \nu (1 + z)^3
\e
while the ionizing flux for a particular species $i$ [one of HI, HeI or HeII] is given by
\be
J_{\nu}(z) \equiv \f{\lambda_i(z;\nu)}{4 \pi} \epsilon_{\nu}(z) 
= \f{\lambda_i(z;\nu)}{4 \pi} \dot{n}_{\nu}(z) h_P \nu (1 + z)^3
\e
The photoheating rate for a particular species is given by
\bear
\Gamma_{PH}(z) = 4 \pi \int_{\nu_{\rm min}}^{\infty} \de \nu
\f{J_{\nu}}{h_P \nu} \sigma_i(\nu) h_P (\nu - \nu_{\rm min})
\nline
= (1 + z)^3 \int_{\nu_{\rm min}}^{\infty} \de \nu ~ \lambda_i(z;\nu) 
\dot{n}_{\nu}(z) \sigma_i(\nu) h_P (\nu - \nu_{\rm min})
\ear
where $\nu_{\rm min}$ is the threshold frequency for the species considered.
The photoionization rate is given by
\bear
\Gamma_{PI}(z) &=& 4 \pi \int_{\nu_{\rm min}}^{\infty} \de \nu
\f{J_{\nu}}{h_P \nu} \sigma_i(\nu) 
\nline
&=&
(1 + z)^3 \int_{\nu_{\rm min}}^{\infty} \de \nu ~ \lambda_i(z;\nu) 
\dot{n}_{\nu}(z) \sigma_i(\nu) 
\ear

Note that we have included the clumping term $R(\Delta_i)$ in the expressions 
(\ref{eq:dEdt}) and (\ref{eq:dXHIdt}). 
As more and more regions of higher densities get ionized, 
the value of 
$R(\Delta_i)$ becomes 
larger which, in turn, gives a larger value of the temperature. Thus 
the temperature $T$ of the ionized regions 
obtained from the above system of equations are essentially
weighted by the mass of the corresponding regions. 
In this sense, one can assume 
$T$ to be an estimate of the mass-averaged temperature of the region. 
We should emphasize that $T$ is {\it not} the rigorously-defined 
mass-averaged temperature, but should be considered as a simple approximation
in the ionized regions.
If the quantities $T$ and $X_i$ defined in this section are 
approximate estimates of the 
mass-averaged values in the ionized region, then the global 
mean values $T_{\rm global}$ and $X_{{\rm global}, i}$ can be obtained by weighted averages over different regions, 
according to the mass fraction of the corresponding region.
Also, note that the above $T$ is {\it not} same as the conventional $T_0$, which is defined as the temperature of gas 
at the mean IGM density ($\Delta = 1$). Similarly, the ionization fractions defined above need not 
correspond to the values at the mean density. The values at the mean density (i.e., $T_0, X_{{\rm HI},0}$) can be solved using the same equations 
(\ref{eq:dTdt}) and (\ref{eq:dXHIdt}), but {\it without} putting in the clumping factor $R(\Delta_{\rm HII})$, i.e.,
\bear
&&\!\!\!\!\!\!\!\!\!
\f{\de T_0}{\de t} = -2 H(z) T_0 - \f{T_0}{\sum_i X_{i,0}} 
\f{\de \sum_i X_{i,0}}{\de t}
\nline
&&\!\!\!\!\!\!\!\!\!
+ \f{2}{3 k_{\rm boltz} n_B} 
\left[n_{{\rm HI},0} \f{\Gamma_{\rm PH}(z)}{Q_{\rm HII}(z)}
-  \alpha_{RC}(T_0) n_{{\rm HII},0} n_{e,0} (1 + z)^3\right]
\nline
\ear
and
\be
\f{\de X_{{\rm HI},0}}{\de t} = 
- X_{{\rm HI},0} \f{\Gamma_{\rm PI}(z)}{Q_{\rm HII}(z)}
+ \alpha_R(T_0) X_{{\rm HII},0} X_{e,0} \f{\bar{\rho}_B}{m_p} (1 + z)^3
\e

\subsection{Mean free path for photons}

The mean free path for ionizing photons depends on the size and topology of the ionized regions. Hence, for a self-consistent calculation 
of the mean free path, one has to use the evolution of the volume filling factor of the ionized regions. Note that we only have statistical information 
about the fraction of volume and mass which is ionized, i.e. we do {\it not} calculate the size of individual ionized regions. 

Given this situation, we use a simple model developed by MHR to calculate the mean free path $\lambda_{i,0}(z)$ for photons (at $\nu = \nu_H$). 
As discussed in MHR, their method is  a good approximation when a very high fraction of volume is ionized. It is clear that a photon will be 
able to travel through the low density ionized volume
\be
F_V(\Delta_i) = \int_0^{\Delta_i} \de \Delta ~ P(\Delta)
\e
before being absorbed. In  the simple model, one assumes that the fraction of volume filled up by the high density regions is $1 - F_V$, hence their size is proportional 
to $(1 - F_V)^{1/3}$, and the separation between them along a random line of sight will be proportional to $(1 - F_V)^{-2/3}$, which, in turn, will determine the 
mean free path. Then one has
\be
\lambda_{i,0}(z) = \f{\lambda_0}{[1 - F_V(\Delta_i)]^{2/3}}
\label{eq:lambda_0}
\e
where we can fix $\lambda_0$ by comparing with low redshift observations. In fact, it has been suggested (from simulations and structure formation arguments; MHR) 
that $\lambda_0$ should be determined by the Jeans length which, in turn, 
is determined by the minimum circular velocity for star-forming haloes 
[equation (\ref{radfeed})]:
\be 
x_b(z) = H_0^{-1} v_c \sqrt{\f{\gamma}{3 \Omega_m (1 + z)}}, 
\e
where $\gamma$ is the adiabatic index.  
In this work, we assume $\lambda_0 \propto x_b(z)$, with the
proportionality constant being determined by comparing with low
redshift observations. The mean free path for photons at $\nu=\nu_H$
is given by the typical separation between the Lyman-limit systems,
which is observed to be $\sim 33$ Mpc at $z = 3$. From the knowledge
of $\lambda_{i,0}(z)$ on can then predict the number of Lyman-limit
systems per unit redshift range through the relation
\cite{mhr99,miralda03}
\be
\f{\de N_{\rm LL}}{\de z}
= \f{c}{\sqrt{\pi}~\lambda_{i,0}(z) H(z) (1+z)}
\e
which can be directly compared with available observations at $2 < z < 4$.

\section{Properties of absorption systems}

In this section, we discuss about how to obtain some of the properties of the IGM when they are observed in the absorption spectra of QSOs. 

Given the probability distribution, the estimates of the temperature, $T$,  and of the neutral hydrogen density $n_{\rm HI}$, we can compute the mean 
transmitted flux as will be observed in the absorption spectra of QSOs. The optical depth at a given point is given by
\be
\tau({\bf x},z) = I_{\alpha} n_{\rm HI}({\bf x},z) (1 + z)^3 \f{c}{H(z)}
\e
where
$I_{\alpha} = 4.45 \times 10^{-18}$ cm$^{-2}$ is a constant.
It is natural to assume that the matter and radiation in the IGM are in photoionization equilibrium; in that case, the neutral hydrogen density is related to the 
baryonic overdensity through
\be
n_{\rm HI}({\bf x},z) = n_{{\rm HI},0}(z) 
\Delta^{2.7 - 0.7 \gamma}({\bf x},z)
\e
The optical depth in the ionized region will then be
\be
\tau({\bf x},z) = A(z)
\Delta^{2.7 - 0.7 \gamma}({\bf x},z)
\e
where
\be
A(z) = I_{\alpha} n_{{\rm HI},0}(z) (1 + z)^3
\f{c}{H(z)}
\e
In general, one should use the global average value of $n_{{\rm HI},0}(z)$  in the above expression, taking into account the neutral and different ionized 
regions. The transmitted flux is
\be
F({\bf x},z) \equiv {\rm e}^{-\tau({\bf x},z)};
\e
with its global mean given by 
\be
F(z) 
= \int_0^{\Delta_{\rm HII}} \de \Delta ~ 
{\rm e}^{- A(z) \Delta^{2.7 - 0.7 \gamma}} ~ P(\Delta)
\e
The equation of state (EOS) can be written in terms of the adiabatic index as  $T\propto \Delta^{\gamma-1}$; note that sometimes, and somewhat confusing, $\gamma$ rather then $\gamma-1$ is defined as the slope of the temperature - density relation. 
It is, in principle, possible to compute the value of $\gamma$ by studying the evolution of the temperature for fluid elements of 
different densities. However, such a study is somewhat beyond the scope of this paper, and will be reported somewhere else. As far as this work
is concerned, we keep $\gamma$ as a free parameter varying in the range $1 \le \gamma \le 2.4$ and compute the transmitted flux over a wide range of values of $\gamma$. Note that the choice of $\gamma$ does not affect any of our {\it other} results.

Another quantity that can be readily estimated from our models is the optical depth of CMB photons due to Thomson scattering 
with free electrons. This can be written as
\be
\tau_{\rm el}(z) = \sigma_T c \int_0^{z[t]} \de t ~ n_{{\rm global}, e} ~ 
(1+z)^3 
\e
where $n_{{\rm global}, e}$ is the global average value of the comoving electron density. We neglect additional small contributions to $\tau_{\rm el}$ arising from early X-ray sources but we do include relic free electrons from cosmic recombination \cite{vgs01}.

\section{Results}

In this section, we present the main results for our model along with their interpretation. In the first part, we analyze what we consider the ``fiducial'' model in terms 
of the choice of free parameters, and confront it with observations. We shall see that this  ``fiducial'' model seems to match all the available experimental data, thus
justifying our analysis. 

The two main free parameters of the model are the ionizing efficiencies of  the two stellar populations, both of which are quite uncertain. For ``normal''
PopII stars, the fiducial values are taken to be $\epsilon_{\rm SF, PopII} = 0.1$, which is partly constrained by low redshift observations of the 
cosmic SFR \cite{nchos04}, 
and $f_{\rm esc} = 0.5\%$, yielding $\epsilon_{\rm PopII} = 0.05\%$.  The estimates of the escape fraction are quite uncertain, and we shall 
discuss the effects of its variation later on. 

For the metal-free PopIII stars, 
the best observational constraints come from the excess in the cosmic 
NIRB data \cite{sf03,msf03}, which imply an upper limit on the combination
$\epsilon_{\rm SF, PopIII} (1 - f_{\rm esc}) 
\equiv \epsilon_{\rm SF, PopIII} - \epsilon_{\rm PopIII} \leq 0.1$ for 
top-heavy IMF. However, this is 
not sufficient to constrain $\epsilon_{\rm PopIII}$.
It is found that for low mass haloes ($M \sim 10^6 M_{\odot}$), 
the escape fraction can be as high as $f_{\rm esc} \approx 0.95$ \cite{wan04}, while 
it can be significantly low for high-mass haloes \cite{rs00,kysu04}.
Since the mass function of collapsed haloes is dominated 
by low-mass haloes at high redshifts, it is reasonable to use 
high values of $f_{\rm esc}$, particularly before reionization. 
Once the IGM is ionized, the star formation in low mass haloes 
is suppressed and hence one should probably use 
a lower value for the escape fraction appropriate for haloes of higher mass.
Similarly there are no strong (theoretical or observational) constraints on the value of
$\epsilon_{\rm SF, PopIII}$.
Hence, for simplicity, we ignore any redshift or mass-dependence of $f_{\rm esc}$ and 
$\epsilon_{\rm SF, PopIII}$ and 
use the fiducial values
$\epsilon_{\rm SF, PopIII} = 0.8\%$ and $f_{\rm esc} = 90\%$ so that
$\epsilon_{\rm PopIII} = 0.7\%$. 
Note that all our results (except the SFR) are sensitive only 
to the combination $\epsilon_{\rm SF, PopIII} f_{\rm esc}$. Hence one can 
vary the individual values of the two parameters keeping the value 
of $\epsilon_{\rm PopIII}$ same, and still obtain identical results
for the reionization history.
For example, low-mass haloes might have $f_{\rm esc} \approx 1$; this implies 
that our fiducial model requires 
a star-forming efficiency of $\epsilon_{\rm SF, PopIII} = 0.7\%$. 
On the other hand, in the ionized regions (where the mass function is dominated 
by higher mass haloes), the escape fraction could be significantly 
lower (say, $f_{\rm esc} \approx 10\%$); consequently, the implied value 
of the star-forming efficiency in our fiducial model 
would be $\epsilon_{\rm SF, PopIII} = 7\%$, which 
is probably higher than the theoretical expectations, but still 
does not violate the upper limit from NIRB data. However, if the escape 
fraction is significantly lower than this, it can be quite difficult 
to match the NIRB data and reionization constraints simultaneously.
The effects of varying $\epsilon_{\rm PopIII}$ will be discussed later. 

Finally, the value of the transition redshift $z_{\rm trans}$ is taken to be 10, which seems to be favoured by NIRB 
observations. The best-fit value  $z_{\rm trans} = 8.8$  \cite{sf03} obtained for a 
burst-like mode of SFR might be somewhat larger when we allow the PopIII SFR to decrease exponentially  even after $z = z_{\rm trans}$. 
As explained, radiative feedback, setting the minimum mass of the star forming haloes in the ionized regions, is implemented in our formalism. 
We shall discuss its effects in detail. 

\begin{figure*}
\begin{center}
\rotatebox{270}{\resizebox{0.9\textwidth}{!}{\includegraphics{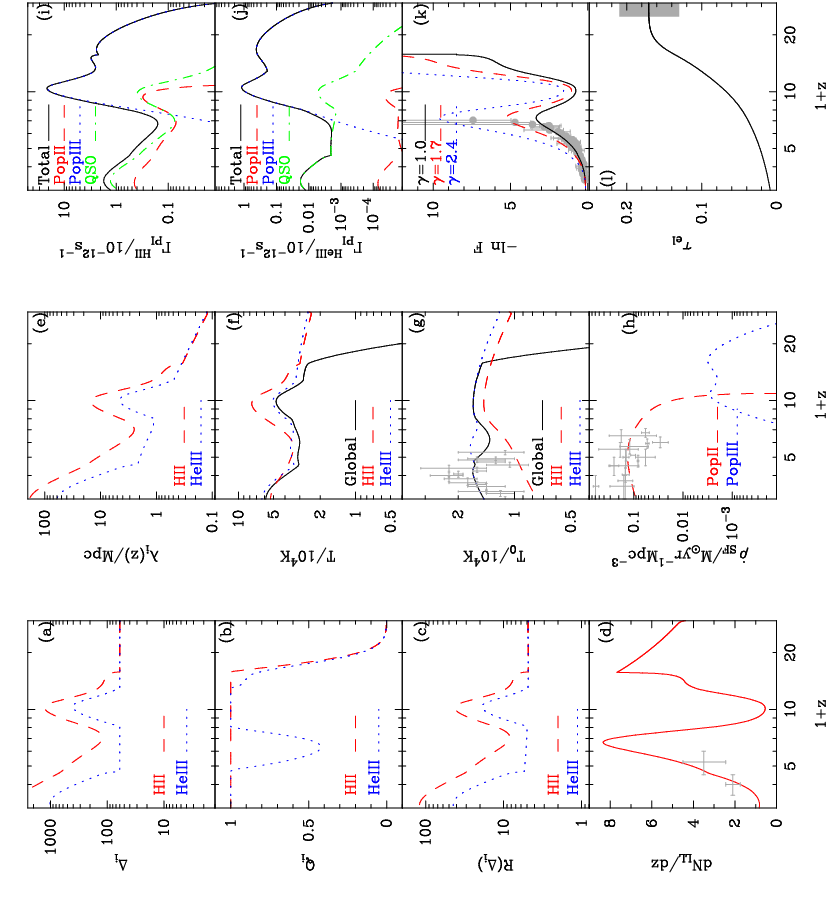}}}
\caption{The fiducial model which matches all available observations. Adopted parameters are $\epsilon_{\rm PopIII}= 7\times 10^{-3}$, $\epsilon_{\rm PopII}=5\times 10^{-4}$,
$z_{\rm trans}=10$.  The Panels show as a function of redshift: (a) critical overdensity for reionization,(b) filling factor of ionized regions, (c) effective clumping factor, (d) specific number of Lyman-limit systems, (e) ionizing photons mean free path, (f) mass-weighted temperature, (g) mean-density gas temperature, (h) cosmic star formation history, (i) photoionization rates for hydrogen, (j) photoionization rates for helium, (k) Gunn-Peterson optical depth, (l) electron scattering optical depth.  See text for detailed description of different Panels.
}
\label{fiducial}
\end{center}
\end{figure*}

\begin{table*}
\caption
{Parameter values used in different figures throughout the paper.}
\begin{center}
\begin{tabular}{|c|c|c|c|c|c|c|}
\hline
Fig. & $\epsilon_{\rm SF,PopII}$ & $f_{\rm esc,PopII}$ & $\epsilon_{\rm SF,PopIII}$ & $f_{\rm esc,PopIII}$ & $z_{\rm trans}$ & $v_c$ \\
\hline
\ref{fiducial}        & 0.1 & 0.005  & 0.007  & 0.9 & 10.0 & using equation (\ref{radfeed})\\
\ref{epsII_lowerlim}  & 0.1 & 0.0005 & 0.007  & 0.9 & 10.0 & using equation (\ref{radfeed})\\
\ref{epsII_upperlim}  & 0.1 & 0.1    & 0.007  & 0.9 & 10.0 & using equation (\ref{radfeed})\\
\ref{epsIII_lowerlim} & 0.1 & 0.005  & 0.002  & 0.9 & 10.0 & using equation (\ref{radfeed})\\
\ref{ztrans_recomb}   & 0.1 & 0.005  & 0.007  & 0.9 & 11.4 & using equation (\ref{radfeed})\\
\ref{vcmin_const_50}  & 0.1 & 0.005  & 0.007  & 0.9 & 10.0 & 50 km s$^{-1}$\\
\hline
\end{tabular}
\label{table:modpar}
\end{center}
\end{table*}

This fiducial model, which seems to fit most of the observations, is shown in Fig. \ref{fiducial}. It also allows us to build a self-consistent
reionization scenario, whose different predictions are compared to the available data in each of the 12 Panels composing the figure.

The reionization history is well exemplified by the evolution of the filling factor of ionized regions $Q_i$ [Panel (b)]. According to the curves shown, hydrogen reionization must have taken place at redshift $z \approx 15$ while 
the HeII reionization is completed around $z \approx 12$. 
We stress here that when $Q_i$ becomes unity, the regions having densities 
less than $\Delta_i$ are completely ionized thus signifying the 
completion of reionization, while regions with higher densities
are completely neutral.
Also note that $Q_{\rm HII} \ge Q_{\rm HeIII}$, thus showing 
that the HeIII ionization front never overtakes the HII front. 
Interestingly, the evolution of $Q_{\rm HeIII}$ 
is markedly affected by the hydrogen reionization; there is a 
decrease in $Q_{\rm HeIII}$ when 
$Q_{\rm HII}$ becomes unity. 
The reason for this is that 
as more and more regions get ionized, the feedback effect 
depresses PopIII star formation, thus decreasing ionization rates. 
However, while this decrease 
is marginal with respect to the wealth of hydrogen ionizing photons, 
it profoundly affects the emissivity of photons above 4 Ryd.  
A more remarkable difference in the evolution of the filling factor 
$Q_{\rm HeIII}$ is however seen at redshifts $3.5 < z < 7$. 
During this epoch the model predicts HeIII recombination followed by  a second  HeII reionization occurring at $z \approx 3.5$; $Q_{\rm HeIII}$  drops to a minimum value of  0.4,
i.e. He is primarily in the singly ionized state at $z\approx 5-6$.  

This behavior can be understood from Panels (i)-(j), where the ionizing rates of the sources responsible for H and He ionization are shown. From these Panels we see that at high redshifts ($>10$), the ionizing flux is totally dominated by PopIII stars, which however fade off at lower redshifts. PopIII stars have a hard spectrum, as seen by comparing the 
values of H and He ionizing rates at high redshifts 
in Panel (i) and (j) respectively. 
It then follows that the PopIII stars can ionize HeII quite efficiently. Once their formation is quenched, there is little source of HeII ionizing photons until the QSOs take over the production of photons above 4 Ryd around redshifts 
of $5$ [see Panel (j)]. Hence while the first ionization of HeII is controlled by PopIII stars, the second one is induced by QSO radiation.  
The transition epoch has been fixed in the fiducial model to 
$z_{\rm trans}=10$, and its effect on the rates can be clearly appreciated; also note that transition from PopIII to PopII stars is {\it not} instantaneous. 
Similar conclusions can be drawn by inspecting the star formation history in Panel (h): PopIII stars produce a first maximum in the star forming activity at $z \approx 15$ where $\dot{\rho}_{\rm SF}\approx 0.003 M_\odot$~yr$^{-1}$~Mpc$^{-3}$, followed by a drop 
because of radiative feedback from ionized regions
and a subsequent raise due to the increasing contribution of PopII stars leading to a less pronounced peak,  $\dot{\rho}_{\rm SF}\approx 0.1 M_\odot$~yr$^{-1}$~Mpc$^{-3}$ at $z=4$. According to these results, only 0.3\% of the stars produced by $z=2$ need to be PopIII stars in order to achieve
the first reionization.

Reionization proceeds from regions of low density to overdense ones, as shown by the evolution of the critical density $\Delta_i$ for the HII and HeIII regions in Panel (a) . We recall that  regions having densities above $\Delta_i$ are neutral, while a fraction $Q_i$ of the regions with densities lower than $\Delta_i$ are ionized. By $z = 5$ ionization fronts have been able to penetrate inside quite dense regions, with $\Delta > 10^3$, leaving therefore tiny islands of neutral gas (mostly in the vicinity or part of galaxies)  in a sea of ionized plasma. As the HII ionization front always leads the HeIII front, it is obvious that the value of $\Delta_{\rm HII}$ is always higher than  $\Delta_{\rm HeIII}$.  Note that $\Delta_i$ is constant as long as $Q_i < 1$.  Panel (c), illustrating the evolution of the clumping factor $R(\Delta_i)$ of the ionized regions, points towards the same behavior. Its evolution is essentially determined by the lognormal distribution, being essentially constant in the pre-overlap stage where $\Delta_i$ is constant. 
The corresponding evolution of the mean free path $\lambda_i(z)$ for H- and HeII-ionizing photons is shown in Panel (e).  

\begin{figure*}
\begin{center}
\rotatebox{270}{\resizebox{0.9\textwidth}{!}{\includegraphics{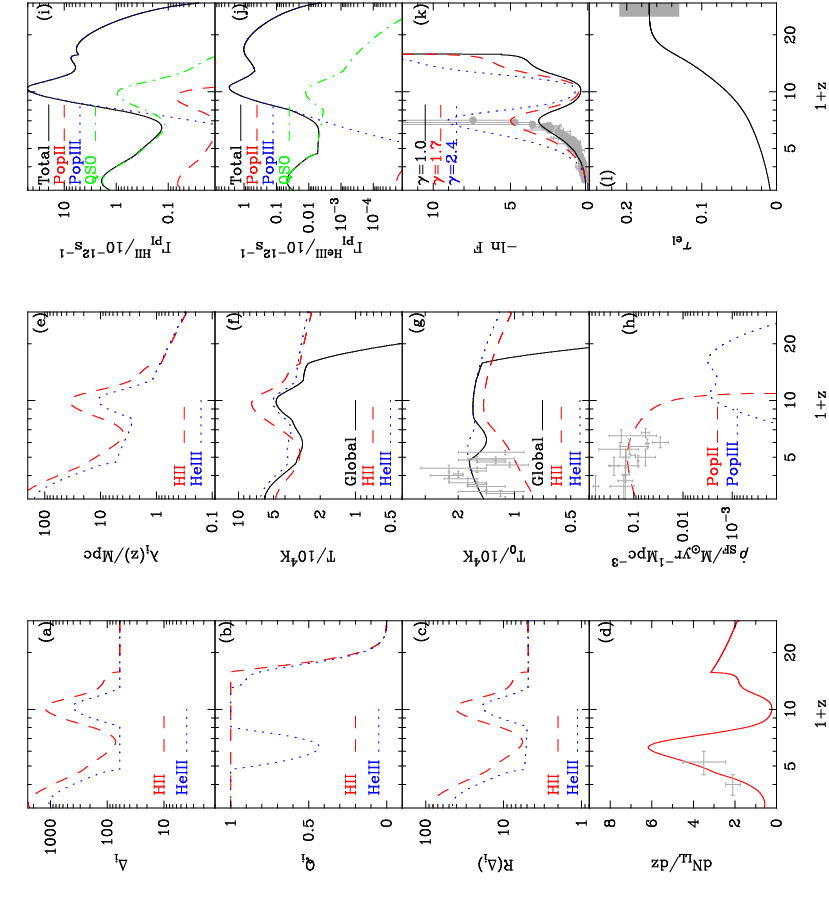}}}
\caption{Same as in Fig. 1, but for $\epsilon_{\rm PopII}= 5 \times 10^{-5}$.
}
\label{epsII_lowerlim}
\end{center}
\end{figure*}

\begin{figure*}
\begin{center}
\rotatebox{270}{\resizebox{0.9\textwidth}{!}{\includegraphics{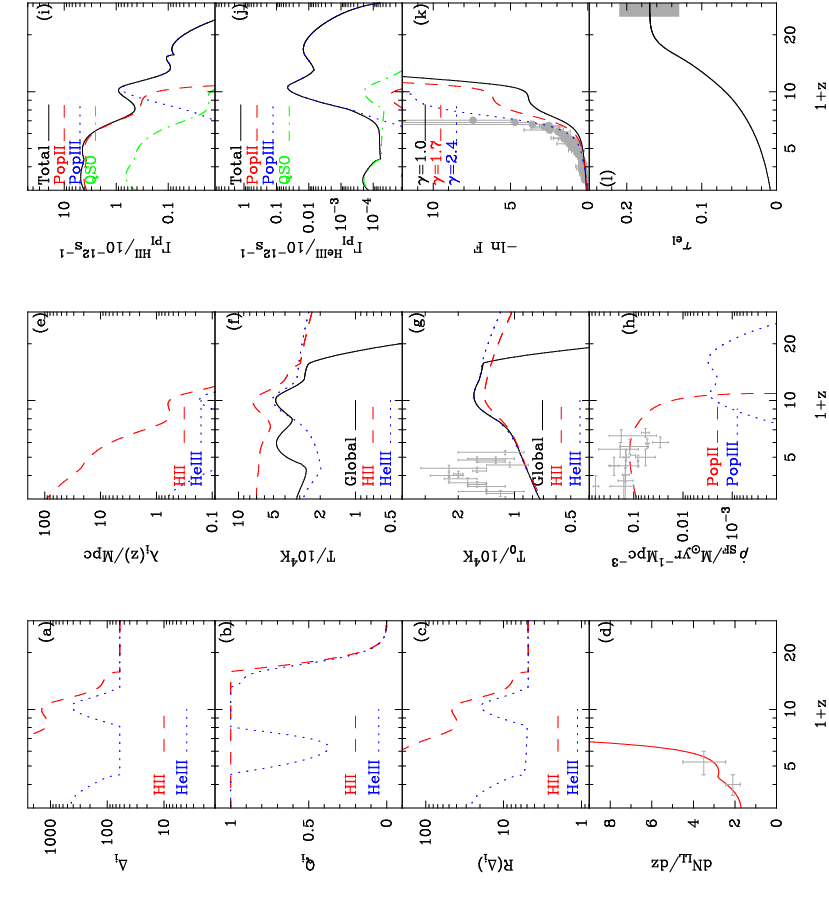}}}
\caption{Same as in Fig. 1, but for $\epsilon_{\rm PopII}= 0.01$
}
\label{epsII_upperlim}
\end{center}
\end{figure*}

\begin{figure*}
\begin{center}
\rotatebox{270}{\resizebox{0.9\textwidth}{!}{\includegraphics{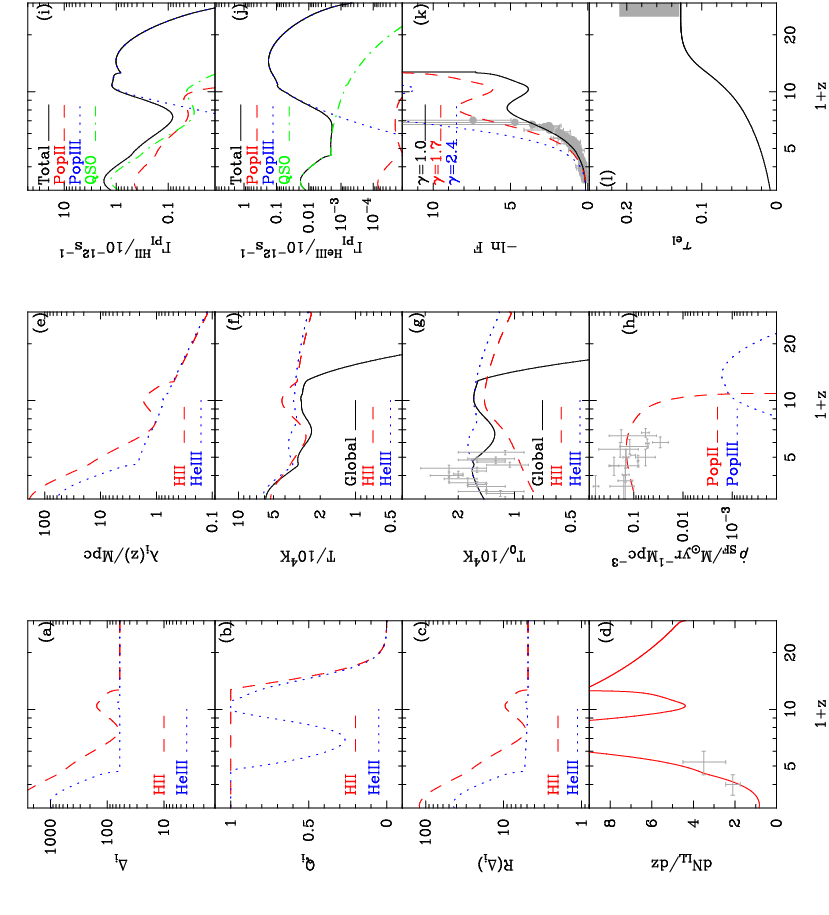}}}
\caption{Same as in Fig. 1, but for $\epsilon_{\rm PopIII}= 0.002$.
}
\label{epsIII_lowerlim}
\end{center}
\end{figure*}

\begin{figure*}
\begin{center}
\rotatebox{270}{\resizebox{0.9\textwidth}{!}{\includegraphics{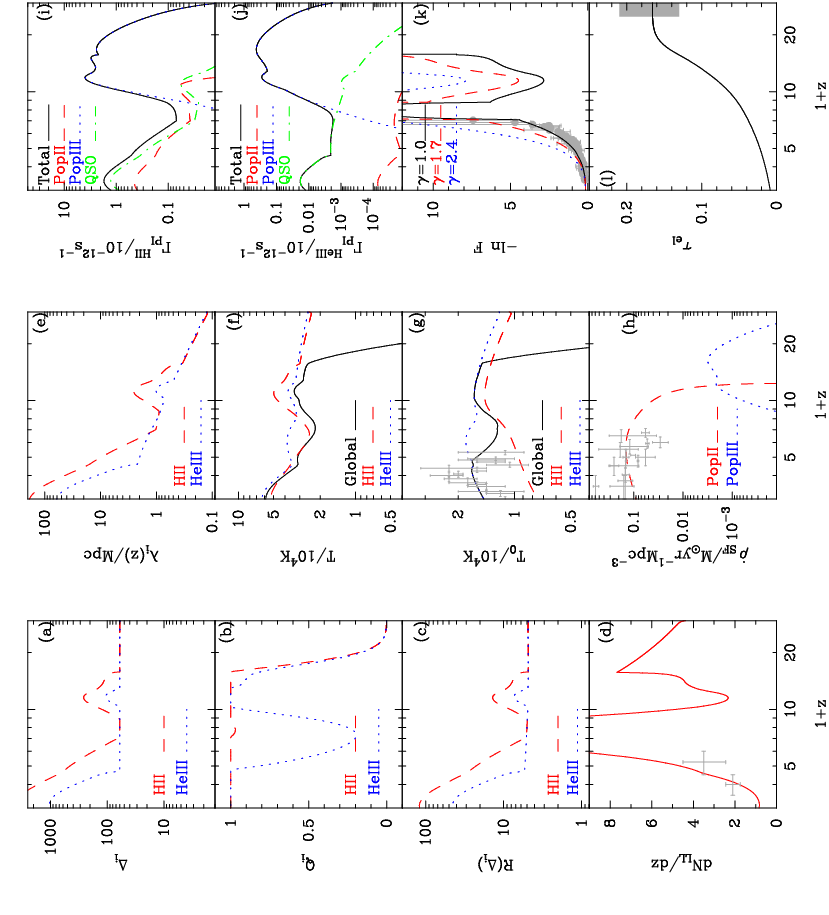}}}
\caption{Same as in Fig. 1, but for $z_{\rm trans}=11.4$. }
\label{ztrans_recomb}
\end{center}
\end{figure*}

\begin{figure*}
\begin{center}
\rotatebox{270}{\resizebox{0.9\textwidth}{!}{\includegraphics{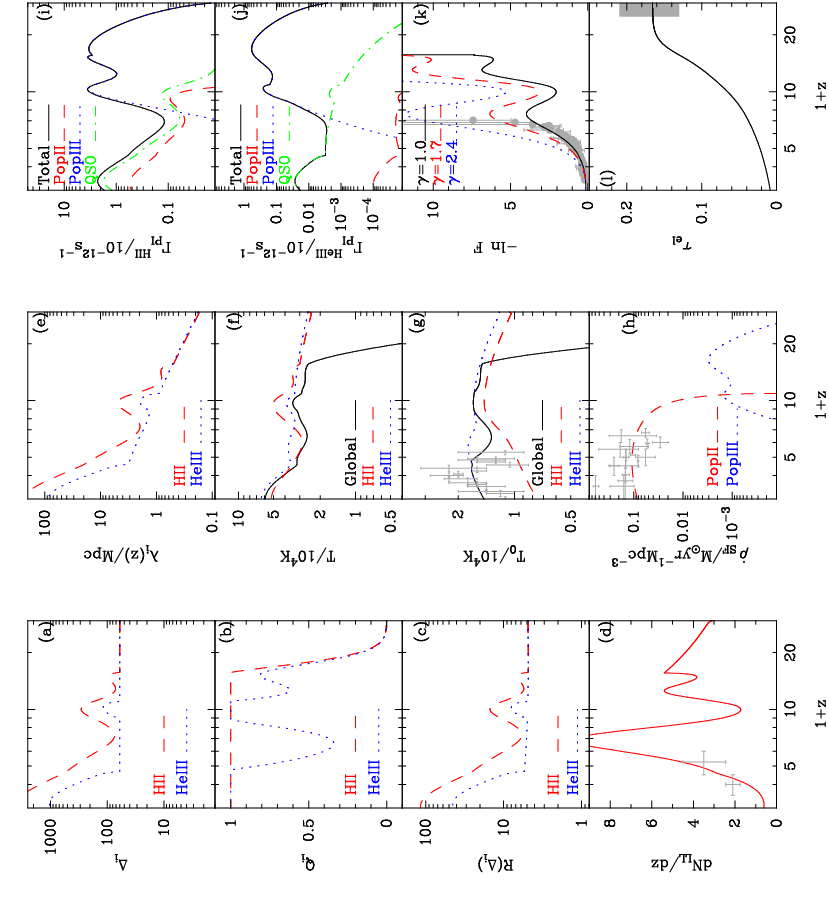}}}
\caption{Same as in Fig. 1, but for a fixed minimum circular velocity of star-forming haloes $v_c =  50$ km s$^{-1}$.}
\label{vcmin_const_50}
\end{center}
\end{figure*}

The derived scenario can now be directly compared with a set of different experimental constraints. 
We have already mentioned that the value of $\epsilon_{\rm SF, PopII}$ is 
chosen such that the predicted SFR agrees with observations. However, 
it is still interesting to note from Panel (h) that our model reproduces the 
observed evolution of the SFR. 
We next consider the number of Lyman-limit systems 
per unit redshift range [Panel (d)], as computed from the mean free path for the hydrogen ionizing photons. 
The data points with error-bars are taken from \citeN{smih94}. 
Another useful comparison involves the mean Ly$\alpha$ transmitted flux, or, 
the effective (Gunn-Peterson) optical depth [Panel (k)], defined as $-\ln F$, where $F$ is the mean transmitted flux. The three curves, from right to left are for $\gamma = 1.0, 1.7, 2.4$, respectively. Although the general raising trend shown by the data \cite{songaila04} is reproduced quite well by the model, there is some indication that $\gamma$ may vary, not unexpectedly, with redshift. In fact, a better agreement with the data is found if  the adiabatic index increases from $\gamma=1$ (isothermal EOS) at $z\approx 3$ to 2.4 at $z=6$. This behavior is opposite to simple expectations based on the assumption that reionization takes place at $z\approx 6$. Although is it true  that at reionization the gas tends to approach isothermality,  afterwards evolving to higher  more adiabatic state, we have to keep in mind that our fiducial  model predicts that hydrogen reionization has occurred long before $z=6$ and therefore it is physically plausible that at later evolutionary stages the EOS deviated from $\gamma=1$; this value is eventually recovered when HeII reionization occurs at $z=3.5$.

As an additional test of the model we next consider the IGM temperature. In particular, Panel (f) shows the estimates of the mass-averaged temperature for HII and HeIII regions and the global one (we recall that this quantity is obtained by weighted averages over different regions, according to the mass fraction of the corresponding region).  Note that this quantity continues to increase with decreasing $z$ at lower redshifts (post-overlap stage) due to the fact that the mass-averaged temperature is dominated by regions of higher densities, and as such higher density regions get ionized gradually, the mass-averaged temperature rises. As already mentioned, the quantity usually derived from QSO absorption line experiments is the temperature at the mean density, $T_0$ [Panel (g)]. Since $T_0$ represents the thermodynamic state of the medium at mean density, its behaviour is markedly different from that of the mass-averaged temperature, particularly in the post-overlap era. As we mentioned before, the mass-averaged temperature rises because regions of higher densities get ionized, while the ionization of such regions do not affect $T_0$ at all. Following the first reionization, $T_0$ is boosted to $1.5 \times 10^4$~K; from there it decreases  because of adiabatic expansion; however, the overall trend is relatively flat. A more pronounced decrease is prevented by the newly available HeII atoms from recombination of this species, which provides extra photoheating to the gas. The curve seems to fit reasonably well the data (taken from from \citeNP{stres00}), although a few of them are more than 1-$\sigma$ away. Apparently, our model would better reproduce the experimental trend if the second HeII reionization could be delayed by roughly 0.5 redshift units.  However, note 
that the data is consistent with the fact that He is singly ionized (HII regions) 
at $z \gtrsim 3.5$, while it is consistent with He being doubly ionized (HeIII regions)
at $z \lesssim 3.5$. Unlike the data points, 
the global temperature in our models rises gradually rather than showing a 
sudden jump. This is related to the fact that $Q_{\rm HeIII}$ has a rather gradual rise.
It might be possible that the data show a sudden jump in $T_0$ because it is based on a few number of lines of sight. Since we expect that there will be large fluctuations in the 
temperature along various lines of sight when $Q_i < 1$, one thus has to compute the temperature along numerous lines of sight to get the global mean, which can then
be compared with the model. On the other hand, our model can be used to generate different lines of sight with different conditions, and can be compared with existing observations.
In this regards, we should mention that there are other analyses of the same data \cite{rgs00} which do {\it not} show the above mentioned sudden jump in $T_0$ around $z \approx 3.5$ mainly because of larger error-bars. In this sense, one requires much robust analysis 
of the observational data for more concrete conclusions. 
Finally, we turn in Panel (l) to the electron scattering optical depth, $\tau _{el}$. The gray shaded region is the 1-$\sigma$ limit as obtained from the 
first year WMAP data. The predicted values match quite well with the WMAP constraint.

We conclude that our fiducial model can explain in a self-consistent manner and simultaneously all the available data existing on cosmic reionization, showing also a strong
predictive power, in spite of its simple and somewhat idealized implementation. The emerging picture is one in which the universe has been initially reionized by massive PopIII stars 
both in H and He; the subsequent disappearance of such exotic stars, required by the NIRB, caused HeIII recombination, followed by its second reionization induced by QSOs. This evolution has produced little effect on hydrogen, which remained essentially ionized throughout, as its ionization state was maintained by normal PopII stars.

In the following sections, we investigate different flavors of such pictures produced by variation of the free unknown model parameters.

\subsection{Constraints on the PopII ionizing efficiency}

The efficiency parameter $\epsilon_{\rm PopII}$ for the low-redshift, normal stars is one of the most ill-determined parameters of our model. We reiterate that we have, mainly for simplicity, assumed the parameter to be independent of $z$ and galaxy mass, although there are hints from the analysis of the {\it Sloan Digital Sky Survey} data that star formation efficiency might vary with the 
halo mass as $M^{2/3}$ 
for galactic stellar masses $< 3\times 10^{10} M_\odot$ \cite{kwh+04}.

As discussed earlier, $\epsilon_{\rm PopII}$ is a the product of the star-forming 
efficiency $\epsilon_{\rm SF, PopII}$ and the escape fraction $f_{\rm esc, PopII}$ of photons from the halo, both of which may vary with redshift and galactic mass. 
The first quantity $\epsilon_{\rm SF, PopII}$ can be reasonably constrained by comparing the model with observations of the SFR at lower redshifts. 
In most cases, it is found to be $\approx 10$\%. The largest uncertainty comes from the escape fraction $f_{\rm esc, PopII}$. It is quite difficult 
to determine it observationally, and the current limits from different kinds of modelling vary from a few per cent to about 50\%. 

In the fiducial model explored so far, we adopted as an educated guess $\epsilon_{\rm PopII}=0.05\%$ while Fig. \ref{epsII_lowerlim} was devised with the aim of determining a lower limit on $\epsilon_{\rm PopII}$ from our models. However, one can realize from there that such determination is quite difficult to make from low ($z < 6$) redshift observations. The reason is that once $\epsilon_{\rm PopII}$ is decreased (in this case by 10 times with respect to the fiducial one) PopII stars play a sub-dominant role and observations can be explained just with the ionizing photons from PopIII stars and QSOs. This does not necessarily mean that the contribution from the PopII stars is insignificant or unnecessary, but it is just that the present observations have very little dependence on PopII star efficiency once below a certain threshold. We believe lower limits on $\epsilon_{\rm PopII}$ should be derived from some other considerations.

On the other hand, it is possible to obtain a stringent upper limit on $\epsilon_{\rm PopII}$, as shown by Fig. \ref{epsII_upperlim}. In fact, if
$\epsilon_{\rm PopII}> 0.01$, then the number of ionizing photons produced will be too high, yielding a too low Ly$\alpha$ optical depth 
(or equivalently, a higher transmitted flux) than the observed value at redshifts around $\sim 3$. Note that for a star forming efficiency of 10\%, this 
upper limit corresponds to an escape fraction of 10\%, which should be considered as quite stringent. 
Also to be noted is that the temperature of the HeIII regions is lower    
than that observed around $z \approx 3.5$, hence this higher value of $\epsilon_{\rm PopII}$ is unable to match the 
high $T_0$ at those redshifts. The reason for this is that with such a 
high production of photons from PopII stars, the ionizing spectrum is no more dominated 
by QSOs and thus effectively becomes 
much softer. This, in turn, results in lower values of $T_0$ for HeIII regions. 

There is one more point which needs to be clarified in this section. We have noted that the change in the value of $\epsilon_{\rm PopII}$ affects the 
low redshift physics only. On the other hand, one can see by comparing Panel (i) and (j)  of Figs \ref{fiducial}, \ref{epsII_lowerlim} and \ref{epsII_upperlim} 
that the values of $\Gamma_{\rm PI}$ are {\it lower} at high redshifts when we increase $\epsilon_{\rm PopII}$. This apparent contradiction 
arises from the fact the value of the photon mean free path $\lambda_0$ in equation (\ref{eq:lambda_0}) is chosen so as to match the low redshift 
observations. When $\epsilon_{\rm PopII}$ is increased, the number of photons is higher, leading to potentially larger ionized regions. In order to control the size of the regions 
so that they match low redshift observations, we have to decrease the value of $\lambda_0$. In turn, this decreases the ionization (and heating) rates at higher redshifts.

\subsection{Constraints on the PopIII ionizing efficiency}

As in the previous section, let us now try to determine the limits on $\epsilon_{\rm PopIII}$ (without changing the value of $z_{\rm trans}$).
In contrast to the case for $\epsilon_{\rm PopII}$, the major uncertainty in $\epsilon_{\rm PopIII}$ comes from the star forming efficiency $\epsilon_{\rm SF, PopIII}$. 
The escape fraction at such high redshifts is usually quite high (nearly 100\%) 
for low mass haloes because typically only a few PopIII stars are required to completely ionize 
the parent galaxy (the masses of of PopIII-hosting galaxies are low). However, our understanding of the star formation process and constraints 
of parameters at high redshifts are quite limited because of the unavailability of direct observations. For the moment, the best constraints on  $\epsilon_{\rm SF, PopIII}$ 
come from the NIRB.

Since $\epsilon_{\rm PopIII}$ does not affect the low redshift observations, the most stringent limits on it come from the constraints on $\tau_{\rm el}$. 
In fact, matching the WMAP constraint requires that this efficiency is confined in the range $0.002 < \epsilon_{\rm PopIII} < 0.03$. 
We would like to stress that, for the 
assumed value of $z_{\rm trans}$, this 
constraint on $\epsilon_{\rm PopIII}$ is independent of the constraints 
obtained from the NIRB data.
The lower limit on $\epsilon_{\rm PopIII}$ implies that in order 
to match the WMAP constraints, one requires a star-forming efficiency of 
$\epsilon_{\rm SF, PopIII} = 0.2\%$ if $f_{\rm esc} = 1$ (probably 
reasonable for low-mass haloes), while in case the escape fraction 
is much lower, say, $f_{\rm esc} = 10\%$ (probably true for 
high-mass haloes), then one needs a star-forming efficiency as 
high as $2\%$.
Also note that the above limits are obtained for a transition redshift  of 
$z_{\rm trans} = 10$. The value of $z_{\rm trans}$ is quite well 
constrained to be $\gtrsim 9$ from NIRB studies. However, for 
the sake of completeness, we would
like to discuss the dependence of these limits on $z_{\rm trans}$.
For higher (lower) values of $\epsilon_{\rm PopIII}$, one has to take higher (lower) values of $z_{\rm trans}$ to obtain the same value of $\tau_{\rm el}$. 
Physically, this implies that one can either let the PopIII stars form efficiently but survive for a shorter time, or let them form inefficiently but survive 
longer, so that the number of free electrons produced is similar. As noted from NIRB studies, the value of $z_{\rm trans}$ we are using is somewhat a 
lower limit. This implies that the lower limit of 0.002 on $\epsilon_{\rm PopIII}$ (the case illustrated in Fig. 4) is quite solid. On the other hand, one can allow for values larger that 0.02 
if the value of $z_{\rm trans}$ is allowed to be larger. 
Thus, given the present observational constraints, 
it is slightly difficult to put a tight upper limit on $\epsilon_{\rm PopIII}$
with uncertainties in $z_{\rm trans}$. On the other hand, if 
we fix the value of $z_{\rm trans}$ from NIRB (or other studies in future), 
our model can constrain the value of $\epsilon_{\rm PopIII}$ quite stringently.

\subsection{Reionization constraints}

In this section we study the constraints on the reionization of 
both hydrogen and HeII with respect to the uncertainties in various
free parameters of the fiducial model. 
We have seen that the fiducial model predicts that HeII was first completely reionized at $z \approx 12$. However, it turns out that 
the values of $z_{\rm trans}$ and $\epsilon_{\rm PopIII}$ can affect the 
HeII reionization at such 
high redshifts. Since HeII reionization is not complete  
before $z \approx 12$, it is obvious that a value of $z_{\rm trans} > 12$
will prohibit a complete reionization of HeII. However, as 
is seen from NIRB studies, it is most likely 
that the value of $z_{\rm trans}$ may not be much larger than 10. 
On the other hand, 
for smaller values of $\epsilon_{\rm PopIII}$, it is 
possible that there are not enough photons from PopIII stars and 
the HeII reionization at high redshifts is {\it not} complete. 
However, in the extreme case where
$\epsilon_{\rm PopIII} = 0.002$ (the lowest allowed value from 
WMAP constraints), we find that 
$Q_{\rm HeIII}$ achieves a maximum value of unity around $z \approx 10$
(see Fig. \ref{epsIII_lowerlim}). 
This implies that HeII must be completely reionized at 
high redshifts as long as the value of $\epsilon_{\rm PopIII}$ does
not violate the WMAP constraint.

Although our fiducial model predicts HeII recombination after its first reionization followed by a second reionization driven by QSOs ionizing power, 
a similar double reionization of H does not appear to have occurred according to our analysis. The physical interpretation of HeII recombination 
at $3.5 < z < 7$ is found in the rather abrupt halt in the PopIII star formation activity at $z \approx 9$; as a consequence, the total number of HeII-ionizing 
photons drops considerably and the second HeII reionization has to wait for the newly available ones produced by QSOs. This scenario seems to 
strongly imply a double HeII reionization (keeping in mind that 
the first reionization might not be complete for lower 
values $\epsilon_{\rm PopIII}$ or for higher values 
of $z_{\rm trans}$).
Hydrogen, however, behaves differently due to the larger availability of photons with energies above 1 Ryd, which are produced also by 
galaxies in addition to QSOs.
Hence the photoionization rate of H remains high enough to keep H atoms in the ionized state. A double H reionization could only occur if the time gap
between the turn-on of PopIII stars and the raise of PopII ones in increased, i.e., if a larger $z_{\rm trans}$ value is assumed. This case has been explored
and is shown in  Fig. \ref{ztrans_recomb}, where we have fixed $z_{\rm trans}=11.4$. Such model predicts a double H reionization around $z=8$, but it is at odd with the constraints from the NIRB, which require PopIII stars down to redshift of about 9 in order to fit the background intensity (and fluctuations) observed in the J band \cite{sf03,msf03}.  
This scenario is a vanilla one for future IGM detection via HI 21-cm line, as the redshifted emission could be observed in the best range of frequencies
and sensitivities of radio telescopes like LOFAR\footnote{http://www.lofar.org} 
\cite{cm03,isfm02}.

\subsection{Different radiative feedback models}

In this section, we analyze the effects of varying the strength of the radiative feedback we have imposed to the reionization process. As we have already noted, the 
radiative feedback affects the HeII reionization as well as star formation rate 
at high redshifts. Hence it is important to check how the results vary once we 
change our assumptions regarding the feedback mechanism.  
In our fiducial model, we have taken the minimum circular velocity of star-forming haloes in the ionized regions to be evolving 
depending on the temperature of the region [see equation (\ref{radfeed})]. However, a different feedback prescription is to fix the
minimum circular velocity of haloes allowed to form stars at a given value, which we take to be $v_c=35$~km s$^{-1}$,
following \cite{gnedin00,ksui01}.

We have first explored such case. However, a comparison with the fiducial model, has shown hardly any significant differences. 
Therefore we have further increased  $v_c$ to 50 km s$^{-1}$ (a value at the upper limit of the physically admissible 
range derived by the above mentioned studies).
The corresponding result  is shown in Fig. \ref{vcmin_const_50}. While most of the results are unaffected, we find that the strong radiative 
feedback suppresses the growth of HeIII regions and delays 
the (first) reionization of HeII.
As one can see from Panel (b), there is substantial decrease in $Q_{\rm HeIII}$
when $Q_{\rm HII}$ becomes unity, thus delaying the HeII reionization 
until $z \approx 10$.
Even this extreme case, however, does not change the qualitative evolution of reionization with respect to the standard one adopted in all
other cases.  Whether this strong radiative feedback has any observational consequence is not clear as all the other considered quantities are essentially unaffected by the choice 
of minimum circular velocity.

\subsection{Reduced power on smaller scales}

Finally, we briefly mention some of the other physical processes which could 
reduce the power of density fluctuations on smaller scales and hence 
influence our results. 
The first such process is the suppression 
of cooling in mini-haloes (i.e., haloes 
having virial temperatures less than $10^4$K) due to photodissociation 
of molecular hydrogen. It is also possible that the minihaloes do not contribute 
to reionization due to confinement of HII regions from feedback effects \cite{rgs02}. 
This would increase effectively the value of $M_{\rm min}(z)$ used in equation (\ref{eq:sfr}). 
Since these minihaloes can potentially contribute 
a substantial number of ionizing photons from PopIII stars 
at high redshifts, it is obvious that one has 
to use a relatively higher value of $\epsilon_{\rm PopIII}$ in the case 
when the minihaloes are not forming stars so as to match 
the WMAP constraints. It turns out that one requires 
$\epsilon_{\rm PopIII} > 0.007$ to match the WMAP limits. 
This implies that if the high mass haloes have an escape fraction  
$f_{\rm esc} = 10\%$, then the required value of star-forming efficiency would be 
as high as $7\%$, while lower values of the escape fraction would 
require higher star-forming efficiencies.
Once such a higher value of $\epsilon_{\rm PopIII}$
is chosen, we found that there is hardly any difference in the reionization 
history of hydrogen when compared to the fiducial model. The 
reionization of HeII at high redshifts, however, occurs much 
earlier because of less severe feedback from ionized HII 
regions; in fact, HeII
reionizes almost simultaneously with hydrogen. To understand this, note 
that the effect of feedback is to suppress the photon-production in 
low mass haloes (which were capable of producing photons before the 
medium was heated up) in the ionized (and heated) regions -- hence it 
should be obvious that the feedback is more severe when there are more
low mass haloes forming stars. It is precisely because of this reason that 
the feedback plays a relatively sub-dominant role when the number of low mass star-forming haloes is reduced. Feedback effects are instead more severe in our fiducial model, where the low mass haloes are allowed to form stars efficiently. 
Interestingly, a similar effect is produced by smaller values of the fluctuation spectrum index $n$, as a result of the reduced small scale power.

\section{Summary and conclusions}

We have developed a simple formalism to study cosmic reionization and the thermal history of the intergalactic medium (IGM). In spite of its simplicity,
the formalism implements most of the relevant physics governing these processes in a self-consistent manner, including the inhomogeneous IGM 
density distribution,  three different classes of ionizing photon sources (massive PopIII stars, PopII stars and QSOs), and radiative feedback inhibiting 
the formation of stars in galaxies below a certain circular velocity threshold. Such approach allows us to predict the star formation/emissivity history
of the source, follow the evolution of H and He reionization and of the intergalactic gas temperature, along with a number of additional predictions
involving directly observable quantities.

By comparing these results with the available experimental data we have selected a ``fiducial'' model. 
This fiducial model self-consistently predicts values matching very well all the available observational data, i.e., the redshift evolution of Lyman-limit absorption systems, Gunn-Peterson and electron scattering optical depths, and the cosmic star formation history without the need to further adjust the free parameters, which are essentially the efficiencies of ionizing photon production for PopIII and PopII stars denoted by $\epsilon_{\rm PopIII}$ and $\epsilon_{\rm PopII}$, respectively. In principle, a third parameter, $z_{\rm trans}$, the epoch of 
cosmic transition from PopIII to PopII stars enters the calculations, but in practice this values is bound to values very close to $z \approx 9$ by the analysis of the Near InfraRed Background (NIRB) data. 

The emerging scenario from our analysis can be summarized in a few points: 
\begin{itemize}
\item 
Hydrogen reionization must have taken place at $z \approx 15$, while 
HeII must have been reionized by $z \approx 12$, taking into account the maximum uncertainty in the value of $\epsilon_{\rm PopIII}$ (we recall that the ionizing photon efficiency is defined as the product of the star formation efficiency and the escape fraction, $\epsilon \equiv \epsilon_{\rm SF} f_{\rm esc}$ for each Population). 
At about $z=7$ HeIII suffered an almost complete recombination ($Q_{\rm HeIII} \approx 0.4$)
as a result of the extinction of PopIII stars (the only contributors of 
photons with energies above 4 Ryd at that redshift), an occurrence 
{\it required} by the interpretation of the NIRB. 
A complete reionization of HeIII occurs at $z=3.5$ and it is driven by hard photons produced by QSOs. Double H reionization does not take  place due to the larger availability of photons above 1 Ryd from PopII stars in galaxies and from QSOs, even after all PopIII stars disappeared. 

\item  Following the first reionization, the temperature of the IGM corresponding to the mean gas density, $T_0$, is boosted to $1.5 \times 10^4$~K; from there it decreases  because of adiabatic expansion; however, the overall trend is relatively flat. Observations are consistent with the predicted temperature of the HII regions at $z \gtrsim 3.5$, while they are consistent with the temperature of the HeIII regions at $z \lesssim 3.5$. This could be interpreted as a signature for the (second) reionization of HeII; however, the global (mass-averaged) temperature rises gradually rather than showing a sudden jump.  This alleged jump in the data
\cite{stres00} might be a spurious feature induced by fluctuations  
along the limited number of lines of sight used by the absorption line experiments, or it 
is possible that the jump does not have a strong enough statistical significance (e.g., see the analysis by \citeNP{rgs00}).

\item PopIII stars produce a first maximum in the star forming activity at $z \approx 15$ where $\dot{\rho}_{\rm SF}\approx 0.003 M_\odot$~yr$^{-1}$~Mpc$^{-3}$, followed by a drop and a subsequent raise due to the increasing contribution of PopII stars leading to a less pronounced peak,  $\dot{\rho}_{\rm SF}\approx 0.1 M_\odot$~yr$^{-1}$~Mpc$^{-3}$ at $z=4$. These results also suggest that only a 0.3\% of the stars produced by $z=2$ need to be PopIII stars in order to achieve the first reionization. We point out that the fiducial model not only correctly predicts at the same time the IGM  reionization and thermal history but it reproduces (with the same free parameters) the cosmic star formation history data at $z < 6$.

In addition to the above features, the data yield the following constraints on these free parameters: $\epsilon_{\rm PopII}< 0.01$, $0.002 < \epsilon_{\rm PopIII} < 0.03$. Varying the efficiencies in these two ranges does not affect most of the general scenario emerging from the first two points above. 
We have experimented with maximally strong radiative feedback finding that the only difference with respect to the fiducial case is to 
suppress the growth of HeIII regions at $z > 10$ and thus delay the  reionization of HeII. Feedback effects impact the ionization history of HeII less severely when the minihaloes (i.e. haloes with virial temperature $T_{\rm vir} < 10^4$~K) at high redshifts are not able to cool and form stars, or when the index of the density power spectrum is smaller. However, 
when the contribution from the minihaloes is ignored, one 
requires a higher value of star-forming efficiency to match the WMAP constraints.

Before we concluding we critically discuss our results in the light of previous numerical works and semi-analytical arguments in the literature. 
A few full numerical simulations of cosmic reionization including radiative transfer have been performed after the WMAP results 
\cite{cfw03,ro04,sahs03,syahs04,gnedin04}. These works use a very different set of assumptions concerning the IMF 
of the first stars, feedback effects and recipes for the sub-grid physics and radiative transfer schemes. Therefore, a detailed comparison is not possible for all these works. \citeN{cfw03} concentrated on reproducing the WMAP Thomson optical depth in a reionization model dominated by ``normal'' (i.e. not very massive)  PopIII stars forming in objects with $T_{\rm vir} > 10^4$~K: minihaloes are thus suppressed (according to our results this should not make a sensible difference) but no radiative feedback is included for objects just above this threshold. This results in a SFR nearly 10 times larger than we predict for $z > 10$.  Because of the compensating effects of IMF and star formation history, hydrogen reionization is achieved at about the same redshift (14.7) as in our case.  It is not clear if this model is able to explain the Gunn-Peterson opacity as the simulation ends at $z\approx 8$. \citeN{gnedin04} took the opposite approach to the problem, focussing on the fit to the Ly$\alpha$ mean transmitted flux. 
The value of $\epsilon_{\rm SF}$ is fixed by normalizing the SFR to its observed value at $z=4$, and therefore, as we also satisfy this constraint,   it should be similar to the one adopted here. The production of ionizing photons (which includes our choice of the IMF and of the escape fraction) is set by construction to the value allowing the best fit to the transmitted flux.  Hydrogen reionization occurs at $z = 6.1\pm 0.3$ in the fiducial model of \citeN{gnedin04} which corresponds to $\tau_{\rm el} = 0.06$ -- quite outside the canonical WMAP value $\tau_{\rm el} = 0.17 \pm 0.04$.  Although the quoted error is somewhat uncertain,\footnote{ 
The uncertainty quoted for this number depends on the analysis technique employed. Fitting the TE cross power spectrum to $\Lambda$CDM models in which all parameters except $\tau_{\rm el}$ assume their best fit values based on 
the TT power spectrum, \citeN{ksb++03} obtain a 68\% confidence range, $0.13<\tau_{\rm el}<0.21$ (the one adopted in this paper). Fitting all parameters simultaneously to the TT and the TE data, \citeN{svp++03} obtain $0.095<\tau_{\rm el}< 0.24$.  Including additional data external to WMAP, these authors were able to shrink their confidence interval to $0.11<\tau_{\rm el}<0.23$.  Finally, by assuming that the {\it observed} TT power spectrum is scattered to
produce the observed TE cross-power spectrum \citeN{ksb++03} infer $0.12<\tau_{\rm el}<0.20$.} 
such a low value for $\tau_{\rm el}$ seems to be very unlikely.  
For this reason Gnedin suggests that the reionization history prior to the hydrogen reionization must have been much more complex than the smooth, monotonic behavior obtained from his simulations. Our results do not require such a complex history. The observed Gunn-Peterson optical depth raise towards $z=6$ is simply caused
by the drop of the photoionization rate [see Panel (i) of Fig. 1] following the disappearance of PopIII stars, thus causing a  significant increase of the hydrogen neutral fraction. To be more quantitative, we find that the mass-averaged neutral fraction 
climbs from $\approx 2\times 10^{-4}$ at $z=4$ to a maximum of $\approx 5\times 10^{-4}$ at $z=6$; following that it decreases until $z=9$ where it reaches the value $\approx 4 \times 10^{-5}$. This behaviour closely tracks (but with opposite derivative)  the evolution of the  photoionization rate in the fiducial model [see Panel (i) of Fig. 1]. It is also consistent with the trend, found e.g. by \citeN{fnswbpr02}, which explains the good agreement with the Gunn-Peterson optical depth of Panel (k). The actual values we find might indicate at $z\approx 6$ a slightly less neutral medium than found by \citeN{fnswbpr02} using Ly$\alpha$ line -- the most reliable determination is $x_{\rm HI} > 5\times 10^{-3}$ when mass-averaged, while $x_{\rm HI} > 2\times 10^{-4}$ when volume-averaged. To compare with the volume-averaged quantity, we can use our predictions corresponding to the mean density, which can be shown to yield the same answer within about  15\%.  For the volume-averaged neutral fraction, we find a value in the range $1-2 \times 10^{-4}$, which is in good agreement with the data. Since \citeN{fnswbpr02} state that the volume-averaged fraction should be considered as more reliable, we can consider this prediction as a success of our model. 

Two more arguments that have been put forward concerning reionization
history are worth discussing. The first point concerns a higher value
of the lower limit (10\%) for the neutral fraction derived by
\citeN{wl04}, based on the size of the ionizing radiation influence
region around two QSOs at $z \approx 6.3$ (see also
\citeNP{mhc04}). In brief, the argument used is that the size of the
ionized region derived from observations is smaller than what expected
if $x_{\rm HI} \la 10^{-3}$, as we argue above. A considerable number
of uncertainties could jeopardize this conclusion:
(a) the lifetime of QSOs; (b) effect of peculiar velocities; (c) radiative transfer (shadowing) effects
occurring in the dense environment surrounding high-redshift QSOs (d) determination of the actual {\rm HII} region size 
from the spectrum. For these reasons we feel that using this argument as a constrain to reionization models is still premature.

The second point concerns the temperature evolution of the
IGM. \citeN{hh03}, following the original proposal by
\citeN{tszktc02}, noted that as long as the universe is reionized
before $z = 10$, and remains highly ionized thereafter, the IGM
reaches an asymptotic thermal state which is too cold compared to
observations at $z=2-4$. This indeed applies to our fiducial case,
which predict a redshift of hydrogen reionization $\approx 15$. There
not only we find that the temperature is decreasing relatively slowly
due to the photoheating provided by He complex reionization history,
but also that global temperature rises gradually (following the smooth
$Q_{\rm HeIII}$ growth) rather than showing a sudden jump. We have
speculated that the observed sudden jump in $T_0$ (i.e. the
temperature of mean density gas) might be a spurious effect due to the
limited available number lines of sight. This hypothesis could be
checked in the future by using our results to generate different lines
of sight to be compared directly with observations. Also, it is 
possible that the statistical 
significance is much weaker, particularly if the error-bars on $T_0$ are 
larger than shown here \cite{rgs00}. 

As a final note, it is useful to remember that our study has not
included one possible additional contribution to the ionizing
background due thermal emission from gas shock heated during cosmic
structure formation, recently suggested by \citeN{mfwb04}. Such
emission is characterized by a hard spectrum extending well beyond 4
Ryd, and according to that study, it is comparable to the QSO
intensity at redshift $\ga 3$. Thermal photons alone could be enough
to produce and sustain He II reionization already at $z = 6$. If this
prediction is correct, the partial recombination of HeIII seen between
redshift 3.5 and 8 in our fiducial model might be prevented by such
radiation. The present results make the test of the He state at these
intermediate redshifts a crucial benchmark to assess the importance of
such thermal emission.

\end{itemize}

\section*{Acknowledgments}

The authors would like to thank R. Salvaterra for useful discussions and
the referee, M. Ricotti, for enlightening comments.

\bibliography{mnrasmnemonic,astropap}
\bibliographystyle{mnras}

\end{document}